\documentclass[nofootinbib,superscriptaddress,aps,prd,showkeys,noshowpacs,onecolumn,10pt]{revtex4-2}
\makeatletter
\def\switch@array{}
\makeatother
\usepackage{array}
\usepackage{eurosym}
\usepackage{graphics}
\usepackage{graphicx}

\usepackage{epsf}
\usepackage{bm}
\usepackage{amsmath,amssymb,amsfonts,mathrsfs,amsthm}
\usepackage{latexsym}
\usepackage{enumerate}
\usepackage{comment}
\usepackage{multirow}
\usepackage{longtable}
\usepackage{tabularx}
\usepackage{booktabs}
\usepackage[dvipsnames,svgnames,x11names]{xcolor}
\usepackage[colorlinks = true,
            linkcolor = Cerulean,
            urlcolor  = magenta,
            citecolor = magenta,
            anchorcolor = NavyBlue]{hyperref}
\usepackage{epstopdf}
\usepackage{bbm}

\def\be{\begin{equation}}
\def\ee{\end{equation}}

\makeatother

\begin{document}

\title{Symmetry-Driven $k$-Essence Cosmological Dynamics}
\author{Andr\'es Lueiza-Colipí}
\email{a.lueiza01@ufromail.cl}
\affiliation{Departamento de Ciencias F\'{\i}sicas, Universidad de La Frontera, Casilla
54-D, 4811186 Temuco, Chile}
\author{Nikolaos Dimakis}
\email{nikolaos.dimakis@ufrontera.cl}
\affiliation{Departamento de Ciencias F\'{\i}sicas, Universidad de La Frontera, Casilla
54-D, 4811186 Temuco, Chile}
\author{Genly Leon}
\email{genly.leon@ucn.cl}
\affiliation{Departamento de Matem\'{a}ticas, Universidad Cat\'{o}lica del Norte, Avenida Angamos 0610, Casilla 1280 Antofagasta, Chile}
\affiliation{Institute of Systems Science, Durban University of Technology, Durban 4000, South Africa}
\affiliation{Centre for Space Research, North-West University, Potchefstroom 2520, South Africa}
\author{Andronikos Paliathanasis}
\email{anpaliat@phys.uoa.gr}
\affiliation{Institute of Systems Science, Durban University of Technology, Durban 4000, South Africa}
\affiliation{Centre for Space Research, North-West University, Potchefstroom 2520, South Africa}
\affiliation{Centro de Investigaci\'on, Innovaci\'on y Creaci\'on (CIIC), Universidad Cat\'olica de Temuco, Temuco, Chile}
\affiliation{Departamento
de Ciencias Matem\'{a}ticas y F\'{\i}sicas,  Facultad de Ingenier\'{\i}a, Universidad Cat\'olica de Temuco, Temuco, Chile}
\affiliation{National Institute for Theoretical and Computational Sciences (NITheCS),
South Africa}

\begin{abstract}
We investigate a family of $k$-essence cosmological models selected by the requirement that the field equations admit a nontrivial variational symmetry. Inside this general family of theories, we concentrate on the simplest modification to quintessence, induced by power-law terms of the kinetic energy. The symmetry generator introduces a conservation law which we use to obtain an exact vacuum solution. We generalize our analysis and investigate the structure of the phase-space of the cosmological field equations by performing a dynamical systems analysis, both in the absence and the presence of a pressureless matter source. We employ Hubble-normalized variables and the Poincar\'{e} compactification, so as to determine the asymptotic behavior of the model and identify all admissible cosmological epochs, from early-time to late-time behavior. We discuss in detail the existence and stability conditions of each fixed point as functions of the free parameters, and characterize the full asymptotic structure of the cosmological history as provided by this symmetry-driven $k$-essence model. We explore the conditions under which the model can describe cosmic acceleration. We find that this $k$-essence model successfully describes the inflationary epoch with a natural exit to the matter-dominated era, with the inflaton ultimately evolving to a dark matter source.
\end{abstract}
\maketitle

\section{Introduction}

It has been proposed that an exotic component of the cosmic fluid, named dark energy, is responsible for the present accelerated expansion of the universe \cite{Riess1998,Schmidt1998,Perlmutter1999}. The cosmological constant, $\Lambda $, is the simplest candidate for dark energy and, despite its well-known theoretical shortcomings \cite{Zeldovich1968,Weinberg1989,Perivolaropoulos2022}, the $\Lambda $CDM model has until recently served as the standard framework for describing late-time cosmic acceleration. However, the persistent tensions in cosmological parameters \cite{CosmoverseTensions}, together with recent observational developments, most notably the Baryon Acoustic Oscillation (BAO) measurements from the Dark Energy Spectroscopic Instrument (DESI), support a dynamical character for
dark energy \cite{DESI2025_DR2}. Alongside dark energy, a variety of independent astrophysical and cosmological observations point to the existence of dark matter, a type of matter that does not interact electromagnetically (or interacts very weakly). Evidence for its existence comes from multiple independent probes: the flat rotation curves of galaxies \cite{Rubin1970,Rubin1980}, gravitational lensing observations, both strong and weak \cite{Clowe2006,Massey2010}, the anisotropies of the cosmic microwave background \cite{Planck2018}, and large-scale structure formation \cite{Davis1985,Eisenstein2005}. Recently, the LUX-ZEPLIN (LZ) experiment provided intriguing results on a possible candidate dark matter event \cite{LZ2026}, whose nature is still under ongoing scientific debate. Together, these independent lines of evidence provide strong support for the existence of a non-baryonic matter content that dominates the total matter budget of the universe.

On large scales, the universe is observed to be approximately isotropic, homogeneous, and spatially flat, described by the Friedmann--Lema\^{\i}tre--Robertson--Walker (FLRW) geometry, which possesses a six-dimensional isometry group. In order to account for these geometric properties, a phase of rapid accelerated expansion in the early universe
known as cosmic inflation has been proposed \cite%
{Starobinsky1980,Guth1981,Linde1982,Albrecht1982,Achucarro2022}. This primordial accelerated epoch is believed to have been far more dramatic than the present dark energy period, such that by the end of inflation, no information regarding the initial state of the universe had survived. Inflation was originally introduced to resolve the flatness and horizon
problems and was later recognized as providing a compelling mechanism for generating the primordial perturbations that describe the universe \cite%
{Mukhanov1981,Guth1982,Bardeen1983,Sasaki1986,Mukhanov1988}.

Scalar field models constitute one of the most widely studied frameworks for describing both dark energy and inflation \cite{dn1,dn2,dn3,dn4}. The simplest configuration, that of a canonical field, drives the expansion of the universe when the self-interacting potential dominates, so as to provide a negative pressure required for accelerated expansion \cite%
{Weinberg1989,Linde1982,Ratra1988,Peebles2003,Caldwell1998,Zlatev1999,Brax1999,Starovinky1998}. Nonetheless, several alternative scalar field models have been proposed. Among them, $k$-essence fields, motivated by high-energy physics, generalize the scalar field Lagrangian to be an arbitrary function of the kinetic term $%
X=-\frac{1}{2}\nabla _{\mu }\phi \nabla ^{\mu }\phi $. In this scenario, the accelerated behavior of the universe is due to the dominance of the kinetic term rather than the potential. This class of models was originally introduced in the context of inflation and later extended to describe dark energy \cite%
{ArmendarizPicon1999,ArmendarizPicon2000,ArmendarizPicon2001}. $k$-essence models can also account for a unified dark sector exhibiting a transient dark matter phase together with a late-time de Sitter attractor corresponding to a dark energy component, or a unification of inflation, dark matter and dark energy \cite{Scherrer2004,Bose2009,Bose2009-2,Ferreira2024}, while also forming the matter sector of other interesting gravitational solutions \cite{Pereira1,Pereira2,Pereira3,Pereira4,Pereira5}.

Despite the success of scalar field cosmology and modified gravity in explaining diverse cosmological phenomena, observations constrain gravitation to be consistent with General Relativity on local scales. This is evidenced by solar system experiments 
\cite{Will2011book,Will2014,Bertotti2003}, binary pulsars \cite%
{Damour1992,Kramer2006,Freire2012}, gravitational wave observations of black hole and neutron-star binaries by the Advanced LIGO/Virgo and KAGRA collaborations  \cite%
{LIGO2016,LIGO2019,LIGO2019-2,LIGO2021,LIGO2025,LIGO2026,LIGO2026-2}, and high-precision laboratory tests \cite%
{Will2014,Adelberger2009,Hamilton2015,Takamoto2020,Tino2021}. Consequently, any deviations from General Relativity must be minuscule at these scales. To simultaneously explain both cosmological and local gravitational
phenomena, a screening mechanism is required such that any deviations from General Relativity on local scales, or in high-density regions, are minimal. Some of these mechanisms are chameleon/symmetron screening \cite%
{Khoury2004,Hinterbichler2010}, the Vainshtein mechanism \cite%
{Vainshtein1972,Babichev2013} and kinetic screening, known as $k$--Mouflage, which is characterized by a derivative self-interaction in scalar-tensor theories \cite{Babichev2013,Babichev2009}. Thus, $k$--essence theory can also account for this $k$--Mouflage mechanism, while at the same time being consistent with current cosmological data \cite%
{Brax2013,Haar2021,Bezares2021}.

The gravitational field is described by nonlinear partial differential equations; however, in the isotropic and homogeneous FLRW limit, the cosmological field equations reduce to a system of ordinary differential equations. Due to their inherent nonlinearity, the scalar field potential, as well as the free functions introduced to modify the gravitational action integral, dramatically affect the structure of the solution space. To systematically explore the latter, dynamical systems analysis provides powerful tools for studying the global dynamics of a given cosmological model, revealing important information about the geometric structure of the phase space. Such methods have been widely applied in many cosmological scenarios, including numerous studies in $k$-essence models \cite{Chakraborty2019,Ferreira2024,Csillag2025,Yang2011,Quiros2025,Chatterjee2022,Jorge2007,Shi2021,Fabio2026}. 

Complementary to this approach, symmetry analysis offers a geometric selection rule for determining the free functions of a given cosmological model. Imposing variational symmetries provides several key advantages. First, for each variational symmetry there exists a corresponding conservation law, which can be employed to reduce the order and dimension of the field equations and reveal their integrability properties \cite{Basilakos2011,Tsamparlis2011,Tsamparlis:2018nyo}. Furthermore, constraining the free functions of the cosmological model through the existence of a higher-order symmetry algebra provides a geometric selection criterion, since the generators of the admissible variational symmetries are
intrinsically determined by the underlying gravitational model \cite{Tsamparlis:2018nyo,Christodoulakis2014,Terzis2016,Rez1,Rez2}. In addition, variational symmetries are essential for
the quantization of minisuperspace models, since they provide quantum operators to be used in conjunction with the Wheeler-DeWitt equation \cite%
{Noeth1,Christodoulakis2014-1,Christodoulakis2014-2,Christodoulakis2014-3,Livine,Tadros}. A quantum cosmological description for a class of $k$-essence models was recently introduced in \cite{Lueiza2026}. 

In this work, we focus on the classical properties of a class of $k$-essence models that admit a variational symmetry. In other words, we utilize a geometric selection rule to constrain the free functions of the theory. With the help of the ensuing conserved charge, we obtain a particular exact solution. We then apply a dynamical systems analysis to reconstruct the phase-space structure, aiming to describe the cosmological history and recover the dynamics of the general solution, both with and without matter.

In Section \ref{sec2}, we present the basic properties and definitions of the $k$-essence model. In Section \ref{sec3}, we discuss the cosmological scenario in the context of an FLRW background geometry and introduce the $k$-essence model that possesses a nontrivial variational symmetry. We demonstrate how the latter leads us to the derivation of a particular exact solution. The global dynamics of the cosmological field equations are presented in Section \ref{sec4}. There, we introduce dimensionless variables within the Hubble-normalization framework and identify the stationary points of the system. Each stationary point corresponds to an asymptotic solution describing a specific epoch of cosmic evolution. Furthermore, compactified variables based on the Poincar\'{e} map are employed, revealing new asymptotic behaviors of the cosmological model. The phase-space analysis is performed both in the absence of matter sources and in the presence of a matter component. In Section \ref{sec6}, we summarize our main results. Finally, we include two appendices, in Appendix \ref{appendix:fixed points radiation}, we list the critical points in the case of radiation, while in Appendix \ref{append2}, we introduce the effect of a nonzero mass term for the $k$-essence field described by an exponential potential.  

\section{$k$-essence Theory}

\label{sec2}

The generic $k$-essence model is a second-order theory of gravity, with a scalar field $\phi $ and a nonlinear kinetic term. The most general gravitational action integral for the theory reads 
\cite{ArmendarizPicon2001} 
\begin{equation}
S=\int \!\!\sqrt{-g}F(R,X,\phi )d^{4}x,  \label{ac.01}
\end{equation}%
in which $F$ is an arbitrary smooth and continuous function of the Ricci scalar $R$, corresponding to the metric tensor $g_{\mu \nu }$, the $k$-essence scalar field $\phi$, and its kinetic term $X$, given by the expression%
\begin{equation}
X=-\frac{1}{2}\nabla _{\mu }\phi \nabla ^{\mu }\phi.  \label{kin}
\end{equation}%
The action integral \eqref{ac.01} incorporates a large family of gravitational models. For example, standard quintessence \cite{Ratra1988,Peebles2003}, where the scalar field  $\phi$ is minimally coupled to gravity; in the Einstein frame the function $F$ is given by the expression%
\begin{equation}
F\left( R,X,\phi \right) =\frac{R}{2}+X-V(\phi ).  \label{qq1}
\end{equation}%
The Brans-Dicke model \cite{Brans1961}, this is a theory defined in the Jordan frame with a non-minimally coupled scalar field, in this case the function $F$ is described by 
\begin{equation}
F\left( R,X,\phi \right) =\frac{\phi R}{2}+\frac{\omega _{BD}}{\phi }X.
\end{equation}%
Another example is the tachyonic field \cite{Gibtach,Gorini2004,Rez3}, where%
\begin{equation}
F\left( R,X,\phi \right) =\frac{R}{2}-V(\phi )\sqrt{1-2X}.
\end{equation}

In the general theory, the dynamical evolution of the gravitational field is governed by the subsequent set of equations \cite{Bahamonde2015} 
\begin{equation}
F_{R}\left( R_{\mu \nu }-\frac{1}{2}g_{\mu \nu }R\right) -\left[ \frac{1}{2}%
g_{\mu \nu }\left( F-RF_{R}\right) +\nabla _{\mu }\nabla _{\nu }\left(
F_{R}\right) -g_{\mu \nu }\nabla _{\kappa }\nabla ^{\kappa }\left(
F_{R}\right) \right] -\frac{1}{2}F_{X}\nabla _{\mu }\phi \nabla _{\nu }\phi
=0,  \label{feqmet}
\end{equation}%
which follow from the variation of the action integral \eqref{ac.01} with respect to the metric tensor. We remark that $F_{R}$, $F_{X}$, and $F_{\phi }$ denote partial derivatives of $F$ with respect to the indicated variables.

Furthermore, the equation of motion for the scalar reads 
\begin{equation}
\nabla _{\kappa }\left( F_{X}\nabla ^{\kappa }\phi \right) +F_{\phi }=0
\label{feqphi}
\end{equation}%
which corresponds to the modified Klein-Gordon equation. Indeed, when we substitute $F$ from \eqref{qq1}, we recover the theory of General Relativity with a quintessence scalar field.

Equations \eqref{feqphi} can be written in the equivalent way%
\begin{equation}
G_{\mu \nu }=\frac{1}{F_{,R}}T_{\mu \nu }^{\left( k\right) } \, , \label{ff1}
\end{equation}%
in which $G_{\mu \nu }$ denotes the Einstein tensor, $\frac{1}{F_{,R}}$ defines the effective varying gravitational constant, due to the coupling with the Ricci scalar, and $T_{\mu \nu }^{\left( k\right) }$ is the effective energy momentum tensor given by the expression 
\begin{equation}
T_{\mu \nu }^{\left( k\right) }=\left[ \frac{1}{2}g_{\mu \nu }\left(
F-RF_{R}\right) +\nabla _{\mu }\nabla _{\nu }\left( F_{R}\right) -g_{\mu \nu
}\nabla _{\kappa }\nabla ^{\kappa }\left( F_{R}\right) \right] +\frac{1}{2}%
F_{X}\nabla _{\mu }\phi \nabla _{\nu }\phi .  \label{enr1}
\end{equation}%
The latter describes the scalar field dynamics and the contribution of the nonlinear function $F$, as a matter content that affects space-time geometry.

In this work, we focus on a theory belonging to the general family
\begin{equation}
F(R,X,\phi )=f_{1}(\phi )R+f_{2}(X,\phi ) ,  \label{model1}
\end{equation}%
where $f_{1}$ and $f_{2}$ are smooth and continuous functions. The field equations \eqref{ff1} reduce to%
\begin{equation}
G_{\mu \nu }=\frac{1}{f_{1}(\phi )}T_{\mu \nu }^{\left( k\right) }
\label{ff2}
\end{equation}%
with the energy momentum tensor $T_{\mu \nu }^{\left( k\right) }$ having the expression
\begin{equation}
T_{\mu \nu }^{\left( k\right) }=\frac{1}{2}\left( f_{2,X}\nabla _{\mu }\phi
\nabla _{\nu }\phi +\frac{1}{2}g_{\mu \nu }f_{2}\right) \,+f_{1,\phi \phi
}\left( \nabla _{\mu }\nabla _{\nu }\phi -g_{\mu \nu }\nabla _{\kappa
}\nabla ^{\kappa }\phi \right).
\end{equation}%
The first parenthesis on the right-hand side of the latter expression describes the contribution for the kinetic and potential terms of the scalar field within the cosmological fluid, while the second one follows from
the nonconstant coupling function $f_{1}(\phi )$. If the theory is defined in the Einstein frame, then $f_{1}(\phi )$ is a constant, while when it is defined in the Jordan frame $f_{1}(\phi )$ is a nonconstant
function.

\section{FLRW $k$-essence Cosmology}

\label{sec3}

According to the cosmological principle, on very large scales, the universe is isotropic and homogeneous, described by FLRW geometry, where for the spatially flat universe, the line element for the background geometry is
expressed as 
\begin{equation}
ds^{2}=-N(t)^{2}dt^{2}+a(t)^{2}\left( dr^{2}+r^{2}d\theta ^{2}+r^{2}\sin
^{2}\theta d\varphi ^{2}\right) ,  \label{lineel}
\end{equation}%
in which $a\left( t\right) $ is the radius of the three-dimensional space, and $N\left( t\right) $ is the lapse function. The lapse $N\left(t\right) $ parametrizes the time variable of the system and without loss of generality can be set to $N=1$ at the level of the equations. However, its presence is essential at the Lagrangian level, and prior to variation, in order to correctly reproduce the constraint equation. 

For the gravitational model \eqref{model1},  we find that the cosmological field equations follow from the variation of the point-like Lagrangian \cite{Lueiza2026}%
\begin{equation}
L=\frac{1}{2N}\left( a^{3}f_{2,X}\dot{\phi}^{2}-12a^{2}f_{1}^{\prime }(\phi )%
\dot{a}\dot{\phi}-12af_{1}(\phi )\dot{a}^{2}\right) +Na^{3}\left( f_{2}(\phi
,X)-Xf_{2,X}\right)\label{eq:minisuper_lag} ,
\end{equation}%
with respect to the variables $\left\{ N,a,\phi \right\} $. The dot denotes the total derivative with respect to the
variable $t$. The gravitational field equations are 
\begin{align}
& 12(f_{1}H^{2}N+f_{1}^{\prime }H\dot{\phi}^{2})-\frac{\dot{\phi}^{2}f_{2,X}%
}{N}+2N\left( f_{2}-Xf_{2,X}\right) =0, \\
& 4f_{1}\left( 2N\dot{H}+3N^{2}H^{2}\right) +4\dot{\phi}\left( 2NH-\frac{%
\dot{N}}{N}\right) f_{1}^{\prime }+\dot{\phi}^{2}\left(
f_{2,X}+4f_{1}^{\prime \prime }\right) +4\ddot{\phi}f_{1}^{\prime }  \notag
\\
& -2N^{2}Xf_{2,X}+2N^{2}f_{2}=0, \\
& 6f_{1}^{\prime }(N\dot{H}+2N^{2}H^{2})-\ddot{\phi}f_{2,X}f_{1}^{\prime }+%
\dot{\phi}\left( \frac{\dot{N}}{N}-3NH\right) f_{2,X}-\dot{\phi}\left( \dot{X%
}f_{2,XX}+\frac{1}{2}\dot{\phi}f_{2,X\phi }\right)   \notag \\
& +N^{2}\left( f_{2,\phi }-Xf_{2,X\phi }\right) =0, \\
& f_{2,XX}\left( 2N^{2}X-\dot{\phi}^{2}\right) =0,
\end{align}%
where $H=\frac{\dot{a}}{Na}$ is the Hubble function.

The gravitational field equations can be written in the equivalent form of
\eqref{ff2},%
\begin{eqnarray}
3H^{2} &=&\frac{1}{f_{1}\left( \phi \right) }\rho ^{\left( k\right) },
\label{eq:hcon} \\
\frac{2}{N}\dot{H}+3H^{2} &=&-\frac{1}{f_{1}\left( \phi \right) }p^{\left(
k\right) },
\end{eqnarray}%
in which $\rho ^{\left( k\right) },$ $p^{\left( k\right) }$ are the energy
density and pressure components for the effective energy momentum tensor $%
T_{\mu \nu }^{\left( k\right) }=\rho ^{\left( k\right) }u_{\mu }u_{\nu
}+p^{\left( k\right) }\left( g_{\mu \nu }+u_{\mu }u_{\nu }\right) $, where $u^{\mu
}$ is the four-velocity of the comoving observer, $u^{\mu }=\frac{1}{N\left( t\right) }\delta
_{t}^{\mu }$, $u^{\mu }u_{\mu }=-1$, and 
\begin{align}
\rho ^{\left( k\right) }=&\frac{1}{2}f_2-Xf_{2,X}+\frac{3H}{N}f_1' \dot \phi\\
p^{\left( k\right) }=&\frac{1}{2}f_2+f_1'\left(\frac{2H}{N}\dot\phi-\frac{\dot N}{N^3} \dot \phi +\frac{\ddot \phi}{N^2}\right)+f_1''\frac{\dot\phi^2}{N^2}.
\end{align}

\subsection{A Symmetry Driven k-Essence Model}

At this point, we further restrict the $k$-essence model at hand so that the action admits, a nontrivial symmetry configuration, and specifically, a scaling symmetry. Theories of the form $f_1=$const. $f_2(\phi,X)= F(X)/\phi^2$ exhibit a scaling symmetry vector \cite{kesymmetry}
\begin{equation} \label{generator0}
  \xi = N \frac{\partial}{\partial N} + \frac{a}{3} \frac{\partial}{\partial a} + \phi \frac{\partial}{\partial \phi}   .
\end{equation}
Considering a smooth function $F(X)$ we may write the series expansion around $X=0$
\begin{equation}
   f_2(\phi,X)=  \frac{1}{\phi^2} \sum_{k=0}^{\infty} c_k X^k,
\end{equation}
where $c_k$ are constants. By introducing a transformation for the scalar field $\phi\rightarrow \tilde{\phi} = \ln \phi$, which also implies $X\rightarrow \tilde{X}=X/\phi^2$,  the above function reduces to
\begin{equation}
   f_2(\tilde{\phi},\tilde{X})= \sum_{k=0}^{\infty} c_k e^{2(k-1)\tilde{\phi}}\tilde{X}^k = c_0 e^{-2\tilde{\phi}} + c_1 \tilde{X} +\ldots c_n e^{2(n-1)\tilde{\phi}}\tilde{X}^n \ldots. \label{generalseries}
\end{equation}
This can be understood as quintessence with exponential potential carrying power-law modifications with respect to the kinetic term. For the rest of this work we concentrate our study to a minimal such modification of the form
\begin{equation}
f_{1}(\phi )=f_{0},\qquad f_{2}(\phi,X)=\alpha X+\beta \exp \left(
(n-1)\phi \right) X^{n},  \label{mod1}
\end{equation}%
where $f_0$, $\alpha$, $\beta$ , $n$ are constant parameters, and for simplicity we have dropped the tildes. This will allow us to study the general dynamics invoked by power-law kinetic terms compatible with theories admitting this symmetry.

The transformed symmetry generator in the variables where the functions of the theory are given by \eqref{mod1} is%
\begin{equation} \label{generator}
  \xi=\frac{N}{2}\partial _{N}+\frac{a}{6}\partial _{a}-X\partial _{X}+\partial_{\phi }.
\end{equation}%
In the appendix \ref{append2} we also study the inclusion of a possible exponential term like the one appearing in \eqref{generalseries}.

As a consequence of the aforementioned symmetry, the gravitational field equations in vacuum possess a conservation law with the corresponding charge being given by
\begin{equation}
I=\frac{a^{3}}{N}\left( \dot{\phi}\left( \alpha +n\, \beta\exp ((n-1)\phi
)X^{n-1}\right) -2f_{0} \frac{\dot{a}}{a}\right).
\end{equation}
We shall use the latter for the derivation of an exact vacuum solution corresponding to the particular value of $I=0$. However, we will further investigate the general dynamics of the full solution space and also include a matter content. 

In the case of the additional dust matter content, it is convenient to define the density parameter for the fluid as $\Omega_m=\frac{\rho_m}{6f_0H^2}$, where $\rho_m$ represents the dust energy density, which can be obtained by means of the continuity equation
\begin{equation*}
\dot{\rho}_{m}+3H\rho _{m}=0.
\end{equation*}%
By directly integrating the latter equation, one obtains $\rho_{m}=\rho _{m0}a^{-3},~$where $\rho _{m0}$ is the today's dust energy density.

For the model \eqref{mod1}, the Hamiltonian constraint \eqref{eq:hcon} with the addition of dust reads
\begin{equation} \label{conwihomega}
6f_{0}H^{2}+\frac{\beta }{2^{n}}(1-2n)\exp \left( (n-1)\phi \right) \dot{\phi%
}^{2n}-\frac{1}{2}\alpha \dot{\phi}^{2}-6f_{0}H^{2}\Omega _{m}=0.
\end{equation}%
Furthermore, the dynamical equations for the Hubble function $H$ and the
scalar field $\phi $ are
\begin{align}
\dot H&= -\frac{\alpha \dot\phi^2+12 f_0 H^2+2^{1-n} \beta  \exp((n-1) \phi )
\dot\phi^{2 n}}{8f_0}, \label{eq:hddot_modelA} \\
\ddot\phi&= -\frac{ 3 \times 2^n   \alpha H  \dot\phi^3+\beta \exp\left((n-1) \phi \right)
\dot\phi^{2 n+1} \left(6 H n+(n-1) (2 n-1) \dot\phi\right)}{2^n \alpha 
 \dot\phi^2+2  n \, \beta \, (2 n-1) \exp\left((n-1) \phi \right)\dot\phi^{2 n}},
\label{eq:klein_gordon_modelA}
\end{align}
where now, without loss of generality, we have considered the lapse function to be $N=1$.

\subsection{Vacuum power-law solution} 

In the vacuum case, where $\Omega_m=0$ in the quadratic constraint Eq. \eqref{conwihomega}, the obtained conserved quantity allows us to define a first order equation $I$=const., complementary to the equations of motion. The resulting system is still quite involved and the solution cannot be expressed in terms of elementary functions, save for the particular case where $I=0$. By demanding that on mass shell the value of the conserved charge is zero, in the cosmic time gauge $N=1$, it is easy to verify that the solution is given in terms of a power-law for the scale factor
\begin{align}
 a(t) & = t^{\sigma}, \\
 \phi(t) & = \frac{1}{n-1} \ln \left[\frac{ 2^{1-n } \alpha (1-3 \sigma ) t^{2(n-1)}}{\beta  \left(n  (3 \sigma -2)+1 \right) }\right],
\end{align}
where the constant $\sigma\neq 1/3$ is associated with the coupling constants $f_0$, $\alpha$ and $n$ through the relation
\begin{equation}
  \sigma = \frac{1}{3} -\frac{1}{6 n } \left(1 \pm \sqrt{1-4 \left(\frac{3\,\alpha }{f_0}-1\right) (n -1) n  } \right).
\end{equation} 
We may note that in the limit $n\rightarrow 0$, we obtain $\sigma\simeq \alpha/f_0$, which implies that the relative coupling of the kinetic term dictates the expansion rate. On the contrary, in the large $n$ limit the power $\sigma$ becomes
\begin{equation}
  \underset{n\rightarrow\infty}{\lim} \sigma = \frac{1}{3} \mp \frac{1}{3} \sqrt{1-\frac{3 \alpha }{f_0}}.
\end{equation}
The system can asymptotically (in terms of a large $n$) approach a stiff fluid solution for $\alpha=f_0/3$, although in order to have exactly $\sigma=1/3$ it can be verified from the equations that the $k$-essence term must be completely eliminated, i.e. $\beta=0$.

We should additionally mention that, in the special case $n=1/2$, the above expressions form the general solution of the field equations. In the following phase-space analysis we exclude this particular value from our considerations and we deal with $n\neq 1/2$ where the general exact solution is untractable. We continue with this analysis in the following section, where we also allow for the possibility of the presence of a dust matter source.

\section{Phase-space analysis of Symmetry driven $k$-essence model}
\label{sec4} 

Within the framework of the Hubble normalization approach, we introduce the dimensionless variables 
\begin{align}
x& =\frac{\sqrt{\alpha }\dot{\phi}}{2\sqrt{3f_{0}}H}, \\
\bar{y}& =\alpha ^{-n}\beta (-1+2n)(6f_{0}H^{2})^{n-1}\exp ((n-1)\phi ).
\end{align}

For a further simplification, we perform the scaling $\bar{y}\to y/x^{2n}$. By incorporating these variables, the Hamiltonian constraint is written in the following way
\begin{equation} \label{condynsys}
1-x^2-y-\Omega_m=0.
\end{equation}
The cosmological field equations are expressed in the equivalent form of
first-order ordinary differential equations
\begin{align}\label{dynsysx}
  x' & = \frac{3 x \left((2 n -1) x^4+(1-2 n ) x^2+\left(n  (2 n -1)+1\right) x^2 y+ k ((3-2 n ) n -1) x y+n (2 n -3 + y) y \right)}{2 (2 n -1) \left(x^2+n \, y\right)} , \\ \label{dynsysy}
  y' & = \frac{3 y \left((2 n -1) x^4 + k (n -1) (2 n -1) x^3+(n  (2 n -1)+1) x^2 y-(1-2 n )^2 x^2-n  (1-y) y \right)}{(2 n -1) \left(x^2+n \, y\right)} ,
\end{align}
where the prime denotes a derivative with respect to the e-foldings $\eta=\ln a$ and $k=2\sqrt{\frac{f_{0}}{3 \alpha }}$ is a constant which is positive. Notice that the system is ill-defined for $n=1/2$ which, as previous mentioned is excluded from the analysis. We additionally leave out the cases $n=0$ and $n=1$ because they correspond to pure quintessence, when we are interested in investigating $k$-essence modification from the basic theory. {It is worth noting that $\beta$ does not affect the dynamical evolution of the system, as it can be absorbed via the field redefinition $\phi \to \phi + \frac{1}{n-1} \ln |\beta|$. Consequently, $\beta$ influences the system only through its sign, which determines whether the phase space is restricted to the upper ($y > 0$) or lower ($y < 0$) half-plane.} In these variables, the equation of state parameter for the total cosmic fluid is expressed as
\begin{eqnarray} \label{weff}
  w=-1-\frac{2}{3}\frac{\dot H}{H^2}=x^2+\frac{y}{2n-1}.
\end{eqnarray}
It is important to note that the dynamical system comprised of \eqref{dynsysx} and \eqref{dynsysy} exhibits a singularity on the curve $x^2+n\, y=0$, which is referred to as an impasse surface. We will initially exclude this set of values, returning to study the dynamics of this curve in section \ref{sec:surface}.

Let us comment here that, in the vacuum case, i.e. $\Omega _{m}=0$, we can algebraically solve the Hamiltonian constraint, Eq. \eqref{condynsys}, to express $y$ as a pure function of $x$, thereby reducing the two-dimensional system to a one-dimensional one. In the logarithmic time $\eta=\ln a$, the dynamical equation for this case is
\begin{equation} \label{singlex}
x^{\prime }=\frac{3 (n-1) x \left(x^2-1\right) (x (k-2 k n+2 (n-1) x)+2 n)%
}{2 (2 n-1) \left((n-1) x^2-n\right)}.
\end{equation}
Finally, the latter is integrated to
\begin{align}
& -\frac{\left(k^2 (2 n-1)-8 n+8\right) \ln \left(2 n \left(-k
   x+x^2+1\right)+x (k-2 x)\right)}{\left(k^2-4\right) (2
   n-1)}\nonumber \\
   & -\frac{2 k \sqrt{16 (n-1) n-k^2 (1-2 n)^2} \tan
   ^{-1}\left(\frac{-2 k n+k+4 (n-1) x}{\sqrt{16 (n-1) n-k^2 (1-2
   n)^2}}\right)}{\left(k^2-4\right) (2 n-1)} \nonumber\\
   & +\frac{2 \ln (1-x)}{(k-2) (2 n-1)}+\frac{2 \ln (x+1)}{-2 k n+k-4 n+2}+2
   \ln (x)=\frac{12 (n-1) \eta }{4 n-2}+c_1
\end{align}
where $c_1$ is an integration constant.

Returning to the generic system described by Eqs. \eqref{dynsysx} and \eqref{dynsysy}, we find that it exhibits eight critical points, two of which lie on the singular curve we mentioned previously, and which we leave outside for the time being. The rest, regular, points can be found in Table \ref{tablepoints},  where they are given in terms of coordinates $(x,y)$.

\begin{table}[htbp]
\centering
\caption{Fixed Points and Stability Conditions of the system (\ref{dynsysx}),(\ref{dynsysy})}
\label{tablepoints}
\renewcommand{\arraystretch}{1.3} 

\begin{tabular}{c|c|c|c|p{4.5cm}|p{3.5cm}}
\hline\hline
\textbf{Point $(x,y)$} & $\boldsymbol{\Omega_m}$ & \textbf{$w$} & \textbf{Classification} & $\boldsymbol{n}$ & $\boldsymbol{k}$ \\
\hline
\hline

\multirow{2}{*}{A: $(-1,0)$} & \multirow{2}{*}{$0$} & \multirow{2}{*}{$1$} 
    & Saddle & $n > 1$ & $k>0$ \\
\cline{4-6}
    & & & Repulsor & $n < 1$ & $k>0$ \\
\hline

\multirow{3}{*}{B: $(1,0)$} & \multirow{3}{*}{$0$} & \multirow{3}{*}{$1$} 
    & \multirow{2}{*}{Saddle} & $n < 1$ & $k > 2$ \\
\cline{5-6}
    & & & & $n > 1$ & $0 < k < 2$ \\
\cline{4-6}
    & & & Repulsor & \multicolumn{2}{c}{Otherwise} \\
\hline

\multirow{6}{*}{\shortstack{C:\\$(\Gamma_+,\, \left(\frac{2}{k}-\Gamma_+)\delta\right)$}} & \multirow{6}{*}{$0$} & \multirow{6}{*}{$-1 + k\Gamma_+$} 
    & \multirow{3}{*}{Attractor} & $n \le -1/2$ & $k > \sqrt{2(1-n)}$ \\
\cline{5-6}
    & & & & $-1/2 < n < 0$ & $k > \frac{4\sqrt{n(n-1)}}{1-2n}$ \\
\cline{5-6}
    & & & & $0 < n < 1$ & $k > 0$ \\
\cline{4-6}
    & & & \multirow{2}{*}{Saddle} & \rule{0pt}{15pt} $\frac{1}{2}-\sqrt{\frac{1}{4-k^2}} < n < \frac{2-k^2}{2}$ & $\sqrt{3} < k < 2$ \\
\cline{5-6}
    & & & & \rule{0pt}{15pt} $n < \frac{2-k^2}{2}$ & $k \ge 2$ \\
\cline{4-6}
    & & & Repulsor & $n > 1$ & \rule{0pt}{16pt} $k > \frac{4\sqrt{n(n-1)}}{2n-1}$ \\
\hline

\multirow{6}{*}{\shortstack{D:\\$\left(\Gamma_-,\, (\frac{2}{k}-\Gamma_-)\delta\right)$}} & \multirow{6}{*}{$0$} & \multirow{6}{*}{$-1 + k\Gamma_-$} 
    & \multirow{2}{*}{Attractor} & $-1/2 < n \le 0$ & $\frac{4\sqrt{n(n-1)}}{1-2n} < k < \sqrt{2(1-n)}$ \\
\cline{5-6}
    & & & & $0 < n < 1$ & $0 < k < \sqrt{2(1-n)}$ \\
\cline{4-6}
    & & & \multirow{3}{*}{Saddle} & $\frac{2-k^2}{2} < n < 1$ & $0 < k \leq \sqrt{3}$ \\
\cline{5-6}
    & & & & \rule{0pt}{15pt} $\frac{1}{2}-\sqrt{\frac{1}{4-k^2}} < n < 1$ & $\sqrt{3} < k < 2$ \\
\cline{5-6}
    & & & & $n > 1$ & $k > 2$ \\
\cline{4-6}
    & & & \multirow{2}{*}{Repulsor} & $n < 1$ & $k > 2$ \\
\cline{5-6}
    & & & & $n > 1$ & $\frac{4\sqrt{n(n-1)}}{2n-1} < k < 2$ \\
\hline

\multirow{2}{*}{E: $(0,1)$} & \multirow{2}{*}{$0$} & \multirow{2}{*}{$\dfrac{1}{2n-1}$} 
    & Saddle & $n < 1$ & $k>0$ \\
\cline{4-6}
    & & & Repulsor & $n > 1$ & $k>0$ \\

\hline

\multirow{11}{*}{\shortstack{F:\\$(\frac{1}{k},\, \frac{1-2n}{k^2})$}} & \multirow{11}{*}{$1 + \dfrac{2(n-1)}{k^2}$} & \multirow{11}{*}{$0$} 
    & \multirow{4}{*}{Attractor} & $\frac{2-k^2}{2} < n \leq \frac{7k^2-16}{2k^2-16}$ & $0 < k < \sqrt{3}$ \\
\cline{5-6}
    & & & & $\frac{7k^2-16}{2k^2-16} \leq n < \frac{2-k^2}{2}$ & $\sqrt{3} < k < 2\sqrt{2}$ \\
\cline{5-6}
    & & & & $n < -3$ & $k = 2\sqrt{2}$ \\
\cline{5-6}
    & & & & $n < \frac{2-k^2}{2} \quad \text{or} \quad n \ge \frac{7k^2-16}{2k^2-16}$ & $k > 2\sqrt{2}$ \\
\cline{4-6}
    & & & \multirow{2}{*}{Saddle} & $-1/2 < n < \frac{2-k^2}{2}$ & $0 < k < \sqrt{3}$ \\
\cline{5-6}
    & & & & $\frac{2-k^2}{2} < n < -1/2$ & $k > \sqrt{3}$ \\
\cline{4-6}
    & & & \multirow{5}{*}{Stable Spiral} & $n > \frac{16-7k^2}{16-2k^2}$ & $0 < k < \sqrt{3}$ \\
\cline{5-6}
    & & & & $n < \frac{16-7k^2}{16-2k^2}$ & $\sqrt{3} < k < 2\sqrt{2}$ \\
\cline{5-6}
    & & & & $-1/2 < n < \frac{16-7k^2}{16-2k^2}$ & $k > 2\sqrt{2}$ \\
\cline{5-6}
    & & & & $ n < -1/2$ & $0<k<\sqrt{3} $ \\
\cline{5-6}
    & & & & $ n > -1/2$ & $\sqrt{3}<k\leq 2\sqrt{2} $ \\
\cline{5-6}
    & & & & $n\neq -1/2$ & $k =\sqrt{3}$ \\
\hline

\end{tabular}
\end{table}

\subsection{Regular points}

The points $A$ and $B$ correspond to stiff matter solutions as the effective equation of state is $w=1$. Near these points, the associated spacetime scale factor evolves as a power-law. For all values of $n$ and $k$, these points are unstable (either saddle or repulsors). Consequently, they may describe early-time behavior of the Universe (if repulsors) or a transient phase (as saddles).

On the other hand, the dynamics of the spacetime representing points $C$ and $D$ depend on the values for $n$ and $k$. In order to simplify the analysis for these points we have introduced the constants
\begin{equation}
  \Gamma_{\pm} = \frac{k(2n-1) \pm \sqrt{k^2 (2 n-1)^2+16 n (1-n)}}{4(n-1)}
\end{equation}
and 
\begin{equation}
  \delta = \frac{k (2 n -1)}{2 (n -1)} .
\end{equation}

\medskip

These points represent a spacetime that evolves with a power-law scale factor ($w\neq -1$). These points are only de Sitter for $n=0$, which is of no interest since that case corresponds to quintessence; however, these points can be seen as a modification of de Sitter space-time. Due to the existence of the square root, points C and D exist under the condition
\begin{equation}
    k^2(2 n -1)^2+16 n  (1-n ) >0 ,
\end{equation} 
which holds for the corresponding intervals of the parameters as they appear in Table \ref{tablepoints}. 

The point $E$ has an equation of state that depends solely on the value of the exponent $n$. For $n=0$ (which corresponds to quintessence), this point behaves as a cosmological constant, leading to an accelerated expansion associated with a de Sitter spacetime. For $ -1<n<\frac{1}{2}$, point $E$ describes an accelerating expansion, while for all other values of $n$, it corresponds to a decelerating universe; in both cases with a power-law type of scale factor. Nevertheless, this point cannot represent the late-time behavior of the universe, since it is not stable; instead, it describes either the early-time evolution (when unstable) or an intermediate stage (when it is a saddle point).

Finally, the point $F$ describes a scaling solution where the scalar field behaves as dark matter. Depending on the value of $n$, the stationary point can be a saddle, an attractor or a stable spiral.

\begin{figure}[tbp]
\centering
\includegraphics[width=1\linewidth]{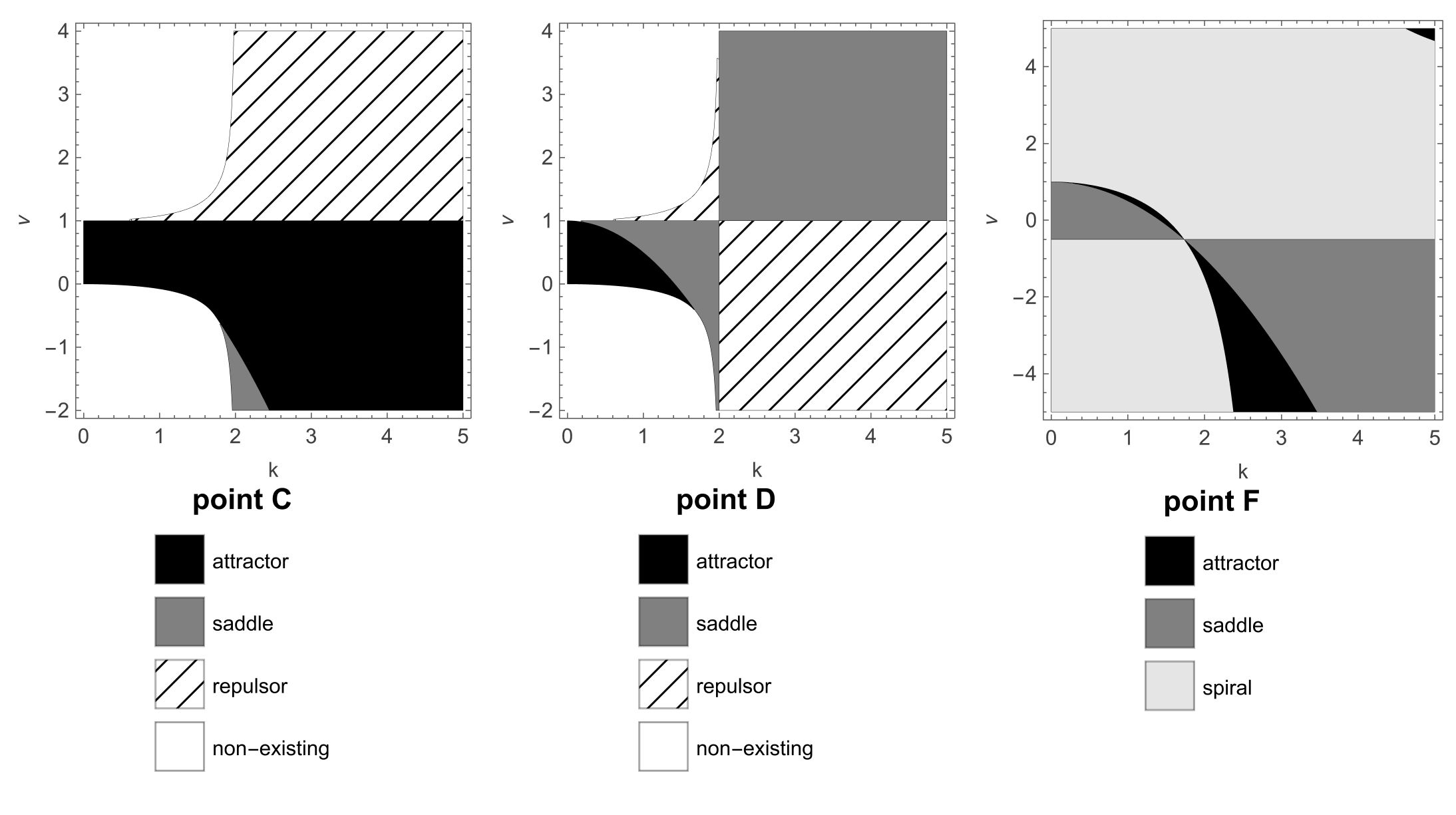}  
\caption{Illustrative depiction of the intervals of the involved parameters affecting the behavior of the critical points C, D and F.}
\label{pointsCDF}
\end{figure}

\begin{figure}[tbp]
\centering
\includegraphics[width=0.7\linewidth]{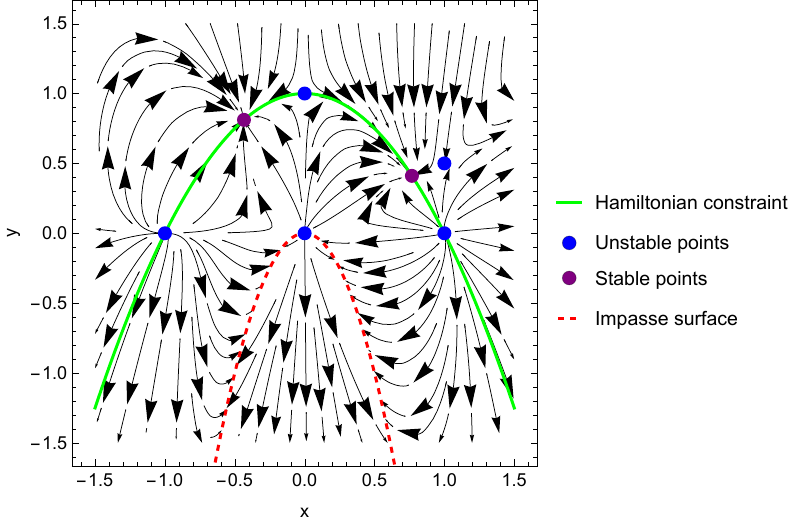}  
\caption{Phase space of the dynamical system with $n=1/4$, and $k=1$. Points 
$A,B,E,F$ are drawn in blue, points $C$ and $D$ are drawn in purple.}
\label{fig:placeholder0}
\end{figure}

\begin{figure}[tbp]
\centering
\includegraphics[width=0.45\linewidth]{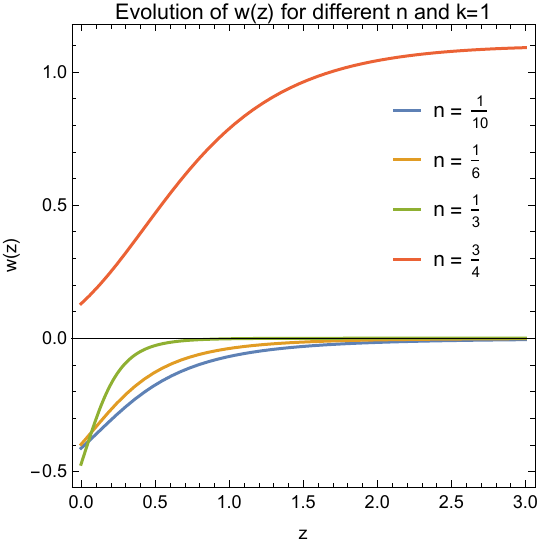}  
\caption{Qualitative evolution of the equation of state parameter for the total fluid $w(z)$  vs redshift $z=\frac{1}{a}-1$ for different values of the index $n$, and $k=1$, with initial conditions $x(0)=0.2$ and $y(0)=0.1$.}
\label{fig:wtot}
\end{figure}

Given the complication of the conditions regarding points C, D and F, we additionally depict in the Fig. \ref{pointsCDF} the corresponding regions in sections of the two parametric space $(n,k)$. In Fig. \ref{fig:placeholder0} we present the phase space portrait that includes the aforementioned points in the particular case $n=1/4$ and $k=1$. In Fig. \ref{fig:wtot} we give the qualitative evolution of the equation of state parameter for the cosmological fluid for different values of the free parameters. 

\subsection{Points at infinity}

Starting from Eqs. \eqref{dynsysx} and \eqref{dynsysy}, we need to compactify the infinity due to the unboundedness of the dynamical system. We perform the compactification in polar coordinates $(\rho,\theta)$, where $\rho\in [0,1]$ and $\theta\in [0,2\pi)$, such that
\begin{equation}
x=\frac{\rho\cos\theta}{\sqrt{1-\rho^2}},\quad y=\frac{\rho\sin\theta}{\sqrt{%
1-\rho^2}}
\end{equation}
with a time reparameterization given by 
\begin{equation}
dT\sqrt{1-\rho^2}=d\eta. \label{eq:time_poinc}
\end{equation}
The fixed points at the infinity $\rho\to 1$ of the dynamical system in these
compactified variables are shown in the Table \ref{tab:critical_points_model_a_poinc_matter}.

\begin{table}[h]
\centering
\setlength{\tabcolsep}{5pt} 
\begin{tabular}{cccccc}
\hline\hline
\textbf{Point} & \textbf{$(\rho,\theta)$} & \textbf{Existence} & $w$ & 
\textbf{Eigenvalues} & \textbf{Stability} \\ \hline\hline
$P_1$ & $(1,0)$ & Always & $\infty $ & $\left\{-3,3/2\right\}$ & 
Saddle \\ 
$P_2$ & $(1,\pi)$ & Always & $\infty$ & $\left\{-3,3/2\right\}$ & 
Saddle \\ 
$P_3$ & $(1,\pi/2)$ & Always & $\frac{\infty}{2n-1}$ & $\left\{0,0%
\right\}$ & Unstable for $n<\frac{1}{2}$, stable otherwise \\ 
$P_4$ & $(1,3\pi/2)$ & Always & $-\frac{\infty}{2n-1}$ & $\left\{0,0%
\right\}$ & Stable for $n<\frac{1}{2}$, unstable otherwise \\ \hline
\end{tabular}%
\caption{Fixed points at infinity with matter.}
\label{tab:critical_points_model_a_poinc_matter}
\end{table}

{At infinity, four equilibrium points appear. The first two, $P_1$ and $P_2$, correspond to ultra stiff singularities, with $w=+\infty$ and because they are always saddle points, they orbits of the system will never reach those points unless a specific fine-tuned condition is given, so $P_1$ and $P_2$ are the locus points of a transient phase of the universe.}

Points $P_3$ and $P_4$ exhibit two vanishing eigenvalues. However, their behavior can be easily deduced by taking the Taylor expansion of the radial equation around $r=1$ at $\theta=\pi/2$ and $\theta=3\pi/2$. To make it more explicit, set $\theta=\pi/2$ in the radial equation together with $r=1-\varepsilon$, where $\varepsilon(t)>0$ corresponds to positive small function. Then, calculate the first non-zero term in the series expansion, we arrive at
\begin{equation}
   \dot{\varepsilon} =\frac{6\sqrt{2}}{1-2n} \varepsilon^{3/2} .
\end{equation}
{This, indicates that for $n<1/2$, $\varepsilon$ grows with time and as a result trajectories in the radial direction get further apart from the infinity point $P_3$. On the other hand, for $n>1/2$, the deviation $\varepsilon$ becomes smaller and $P_3$ transitions to an attractor. A similar calculation shows that $P_4$ has the exact opposite behavior.}

\begin{figure}[tbp]
\centering
\includegraphics[width=0.6\linewidth]{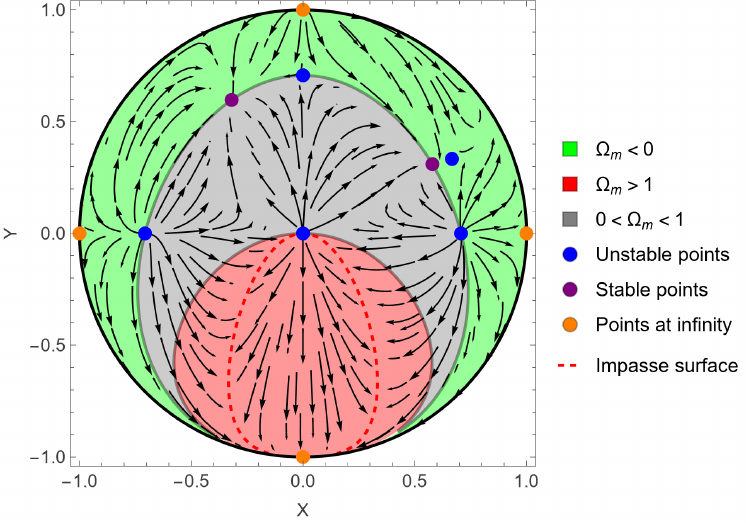}  
\caption{Phase space of the dynamical system in Poincare variables, with $n=1/4$, and $k=1$. We have defined the variables $X=\protect%
\rho\cos\protect\theta,\ Y=\protect\rho\sin\protect\theta$. Points $A,B,E,F$ are drawn in blue, points $C$ and $D$ are drawn in purple, points $P_1,P_2,P_3,P_4$ are drawn in orange. It is worth noting that for $n<\frac{1}{2}$, the only attractor point for $Y<0$ is $P_4$, which is a big rip singularity. While for $Y>0$ the attractor points corresponds to the points $C$ and $D$.}
\label{fig:poinc_disc}
\end{figure}

Fig. ~\ref{fig:poinc_disc} displays the phase-space portrait projected onto the Poincaré disc. Notably, trajectories generated from arbitrary initial conditions never cross the Hamiltonian constraint boundary $\Omega_m = 0$. Since the region with $\Omega_m < 0$ (green) is unphysical, this boundary serves as a natural domain limit without presenting physical inconsistencies. Conversely, the system can dynamically evolve into the $\Omega_m > 1$ region (red), which corresponds to a negative $k$-essence energy density. To restrict the phase space to physically meaningful trajectories, a specific sign for $\beta$ must be imposed, guaranteeing that the allowed phase-space is confined to the upper domain. Furthermore, although the impasse surface (dashed red line) exhibits attractive behavior in Fig.~\ref{fig:poinc_disc}, its stability is not universal; it can act as either an attractor or a repeller depending on the free parameters $(n, k)$ and the direction from which it is approached.

\subsection{The surface $x^{2}+n\, y=0$}
\label{sec:surface}

At this point we return to the singularity curve mentioned at the beginning of the previous section, which we momentarily excluded from our considerations. As we mentioned, the dynamical system presents a singular surface when the denominator of Eqs. \eqref{dynsysx} and \eqref{dynsysy} vanishes, i.e., $n\, y+x^{2}=0$. Surfaces of this kind are referred to as \textit{impasse surfaces} in the mathematical literature on singular (or implicit) differential equations \cite{Chua1989,Chua1989-2,Sotomayor2001}, and as \textit{degenerate surfaces} in the theoretical physics literature \cite{QGS, Saavedra2001, Alexandre2022}. Trajectories can reach such surfaces in a finite interval of the evolution parameter. On the impasse surface, some dynamical equations are transformed into constraints, and as a result certain dynamical variables cease to be independently determined by the equations of motion and become gauge parameters. It is important to emphasize that these surfaces do not, in general, coincide with sets of fixed points of the dynamical system (although isolated fixed points may lie on them); consequently, the attractive or repulsive character of the surface as a whole cannot be assessed through the usual linear stability analysis around hyperbolic fixed points. We first study the fixed points that lie on this surface, and subsequently characterize the behavior of the flow in the vicinity of the surface.

We introduce a new time reparameterization $\sigma$, such that $\frac{d\eta }{d\sigma }%
=x^{2}+n\, y$, which absorbs the vanishing denominator of Eqs. \eqref{dynsysx} and \eqref{dynsysy} yielding a regular vector field. The (regular)  two-dimensional dynamical system becomes

\bigskip 

\begin{align}
\frac{dx}{d\sigma }& =\frac{3 x \left(y (x (k (3-2 n) n-k+n (2 n-1) x+x)+n (2 n-3))+(2 n-1) \left(x^2-1\right) x^2+n y^2\right)}{2 (2 n-1)},
\label{eq:dynamical system impasse 1}\\
\frac{dy}{d\sigma }& =\frac{3 y \left(k \left(2 n^2-3 n+1\right) x^3+x^2 \left(\left(2 n^2-n+1\right) y-(1-2 n)^2\right)+(2 n-1) x^4+n (y-1) y\right)}{2 n-1} .
\label{eq:dynamical system impasse 2}
\end{align}%

As mentioned before, the impasse surface corresponds to $ny+x^{2}=0$. On this surface there are two stationary points, $S_1$ and $S_2$, of the regularized system, which coincide with genuine fixed points of the original dynamical equations, i.e., points at which both numerator and denominator of of Eqs. \eqref{dynsysx} and \eqref{dynsysy} vanish simultaneously. Their cosmological properties are summarized in Table~\ref{tab:critical_points_model_a_surface}.

\begin{table}[h]
\centering
\setlength{\tabcolsep}{5pt} 
\begin{tabular}{cccccc}
\hline\hline
\textbf{Point} & $(x,y)$ & \textbf{Existence} & $w$ & $\Omega_m$ & Stability
\\ \hline\hline
$S_1$ & $(0,0)$ & Always & $0$ & 1 & Unstable \\ 
$S_2$ & $(\frac{4n}{(2n-1)k},-\frac{16n}{(1-2n)^2k^2})$ & Always & $\frac{16 (n-1) n (2 n+1)}{k^2 (2 n-1)^3}$ &  $1-\frac{16 (n-1) n}{k^2 (1-2 n)^2}$ & Saddle/``Center''
\\ \hline
\end{tabular}%
\caption{Fixed points on the impasse surface. For point $S_2$ we use the terminology ``center'' in a loose sense since, due to the impasse curve, the directions of time changes left and right from the latter (see Fig. \ref{S2center}).}
\label{tab:critical_points_model_a_surface}
\end{table}

To determine the stability of the fixed points in the original $\eta$-time (the e-foldings), it is important to track the sign of the factor $ny +x^{2}$, since $d\eta/d\sigma$ can change sign when passing from one side of the curve to the other. When this factor is negative, the direction of the flow in $\sigma$-time is reversed with respect to the original $\eta$-time, and the stability character of the corresponding fixed point must be interpreted accordingly (e.g., a sink in $\sigma$-time corresponds to a source in $\eta$-time, and vice versa).

The point $S_1$ corresponds to a matter-dominated point, so the energy density of the scalar field is zero. The stability of the point is not so simple to determine, because it is a non-hyperbolic fixed point of Eqs.  (\ref{eq:dynamical system impasse 1}) and (\ref{eq:dynamical system impasse 2}), so the usual linear stability analysis cannot be performed. The technique employed to determine its stability properties is the \emph{quasi-homogeneous blow-up} \cite{Seidenberg1968,Dumortier1977,PDE,Alvarez2011,Jardon2019}, commonly used to desingularize non-hyperbolic critical points in dynamical systems.

We perform a change of coordinates resembling a weighted polar representation, with another time-reparametrization $\tau$ which regularizes the system at $r=0$ in the weighted polar coordinates

\begin{eqnarray}
    x=r\cos\varphi,\quad y=r^{2}\sin\varphi, \quad  \frac{d\sigma}{d\tau}=r^{2} \, .
\end{eqnarray}

In this desingularized system, the origin is replaced by the invariant circle $r=0$, $\varphi\in[0,2\pi)$, and the original degenerate point $S_1$ is recovered in the limit $r\to 0$ along each direction $\varphi$. The fixed points on this circle then correspond to different directions of approach to (or departure from) $S_1$ in the original phase space. The fixed points in the \textit{blow-up} are presented in the Table \ref{tab:critical_points_blow_up}.

\begin{table}[htbp]
\centering
\caption{Fixed points in the blow-up at the point $S_1$ in the $\tau$-time, where $\chi_\pm=\arcsin\left(\frac{n\pm\sqrt{n (17 n-16)+4}}{2-4 n}\right)$. The eigenvalues of the points are ordered such that the first one is in the radial coordinate, and the second one is in the angular coordinate.}
\label{tab:critical_points_blow_up}
\renewcommand{\arraystretch}{1.3} 

\begin{tabular}{c|c|c|c|c|c}
\hline\hline
\textbf{Point} & $(r,\varphi)$ & \textbf{Existence} & \textbf{Eigenvalues} & \textbf{Classification} & \textbf{Condition} \\
\hline
\hline

\multirow{2}{*}{$U_1$} & \multirow{2}{*}{$(0,0)$} & \multirow{2}{*}{Always} &\multirow{2}{*}{$\left\{-\frac{3}{2},-6 (n-1)\right\}$}  & Saddle & $n <1$\\
\cline{5-6}
    & & & & Attractor & $n > 1$ \\
\hline

\multirow{3}{*}{$U_2$} & \multirow{3}{*}{$(0,\frac{\pi}{2})$} & \multirow{3}{*}{Always} &\multirow{3}{*}{$\left\{\frac{3 n}{2-4 n},\frac{3 (n-1) n}{2 n-1}\right\}$ }  & Attractor & $n <0\quad \text{or}\quad \frac{1}{2}<n<1$  \\
\cline{5-6}
    & & & & Repulsor & $0<n<\frac{1}{2}$  \\
\cline{5-6}
    & & & & Saddle & $n>1$  \\
\hline

\multirow{2}{*}{$U_3$} & \multirow{2}{*}{$(0,\pi)$} & \multirow{2}{*}{Always} &\multirow{2}{*}{$\left\{-\frac{3}{2},-6 (n-1)\right\}$}  & Saddle & $n <1$  \\
\cline{5-6}
    & & & & Attractor & $n > 1$  \\
\hline

\multirow{3}{*}{$U_4$} & \multirow{3}{*}{$(0,\frac{3\pi}{2})$} & \multirow{3}{*}{Always} &\multirow{3}{*}{$\left\{\frac{3 n}{4 n-2},-\frac{3 (n-1) n}{2 n-1}\right\}$ }  & Repulsor & $n <0\quad \text{or}\quad \frac{1}{2}<n<1$ \\
\cline{5-6}
    & & & & Attractor & $0<n<\frac{1}{2}$ \\
\cline{5-6}
    & & & & Saddle & $n>1$ \\
\hline

\raisebox{-0.4em}{$U_5$} & \raisebox{-0.4em}{$(0,\pi+\chi_+)$} & \raisebox{-0.4em}{$n<0$} & \raisebox{-0.4em}{$\left\{\frac{3n(n-1)\sin\chi_+}{1-2n},\ \frac{6n(n-1)\sin\chi_+}{2n-1}\right\}$} & \raisebox{-0.4em}{Saddle} & \raisebox{-0.4em}{Always} \\[0.8em]
\hline

\raisebox{-0.4em}{$U_6$} & \raisebox{-0.4em}{$(0,-\chi_+)$} & \raisebox{-0.4em}{$n<0$} & \raisebox{-0.4em}{$\left\{\frac{3n(n-1)\sin\chi_+}{1-2n},\ \frac{6n(n-1)\sin\chi_+}{2n-1}\right\}$} & \raisebox{-0.4em}{Saddle} & \raisebox{-0.4em}{Always} \\[0.8em]
\hline

\raisebox{-0.4em}{$U_7$} & \raisebox{-0.4em}{$(0,\pi+\chi_-)$} & \raisebox{-0.4em}{$n>0$} & \raisebox{-0.4em}{$\left\{\frac{3n(n-1)\sin\chi_-}{1-2n},\ \frac{6n(n-1)\sin\chi_-}{2n-1}\right\}$} & \raisebox{-0.4em}{Saddle} & \raisebox{-0.4em}{Always} \\[0.8em]
\hline

\raisebox{-0.4em}{$U_8$} & \raisebox{-0.4em}{$(0,-\chi_-)$} & \raisebox{-0.4em}{$n>0$} & \raisebox{-0.4em}{$\left\{\frac{3n(n-1)\sin\chi_-}{1-2n},\ \frac{6n(n-1)\sin\chi_-}{2n-1}\right\}$} & \raisebox{-0.4em}{Saddle} & \raisebox{-0.4em}{Always} \\[0.8em]
\hline

\end{tabular}
\end{table}

Once the fixed points in the blow-up are computed, we need to perform a \textit{blow-down}, which is to redo the change of coordinates to the original $(x,y)$ variables. The fixed points in the blow-up will correspond to characteristic directions on the blow-down, where the qualitative behavior of the phase space is divided into different regions, separated by curves of constant $\varphi$; near these separation lines, the phase space behaves similarly to the stability properties of the blow-up (keeping in mind possible negative values of $ny+x^2$, which inverts the stability properties). With this in mind, it is now clear that $S_1$ is always an unstable point, in the sense that it can only be stable for certain directions in specific values of $n$.

\begin{figure}
    \centering
    \includegraphics[width=0.6\linewidth]{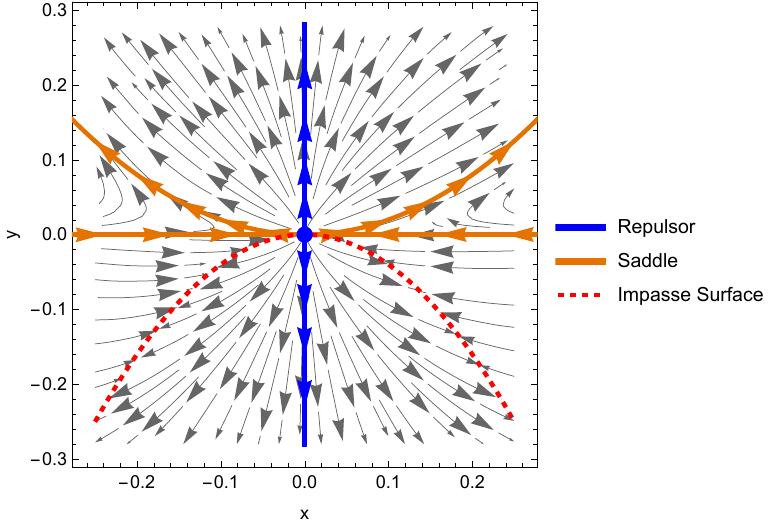}
    \caption{Phase space portrait of the dynamical system \eqref{dynsysx} and \eqref{dynsysy} near the point $S_1$, considering the values $n=1/4$, $k=1$. The arrowed lines portray the blow-up directions; the horizontal lines correspond to $U_1$ and $U_3$, the vertical lines are $U_2$ and $U_4$, and the curved lines are a parabola that corresponds to $U_7$ and $U_8$. The red dashed line is the impasse surface.}
    \label{fig:blowup 1}
\end{figure}

The phase space portrait for the equations \eqref{dynsysx} and \eqref{dynsysy} near the point $S_1$ is portrayed in Fig. \ref{fig:blowup 1}; it is clear that, for $n=1/4,\,k=1$, none of the directions around $S_1$ is stable. It's worth noting that even though the direction $U_4$ corresponds to an attractor in the regularized system for the parameters presented before, in the original (singular) system it is unstable, because in that region the $\sigma$-time changes sign with respect to $\eta$ due to $y$ being negative.

\begin{figure}
    \centering
    \includegraphics[width=0.5\linewidth]{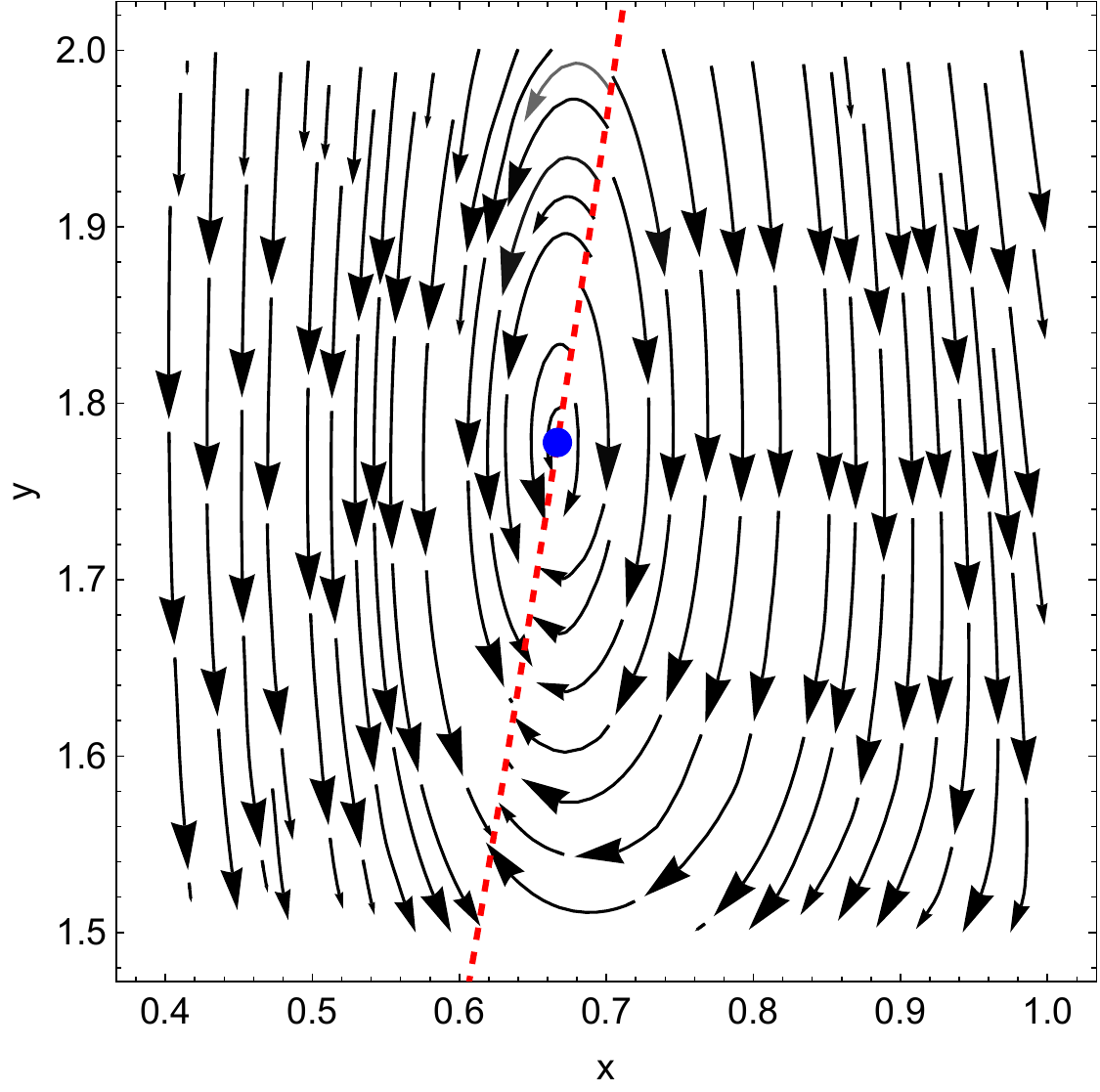}
    \caption{The point $S_2$ for values of the parameters for which it is a center. In this case, $n=-1/4$, $k=1$. The red dashed line represents the impasse curve $y=4x^2$. For these values of the parameters $k$ and $n$, the point $S_2$ corresponds to an accelerating solution $w=-20/27$. The trajectories around this point start and end at the singularity surface.}
    \label{S2center}
\end{figure}

The stationary point $S_2$ describes a solution where the scalar field and the
matter term coexist, that is, $\Omega _{m}\left( S\right) =1-\frac{16}{k^{2}}%
\frac{n\left( n-1\right) }{\left( 1-2n\right) ^{2}}$. Interestingly enough this point can either be a saddle or a center, since its two eigenvalues are
\begin{equation}
  \lambda_{S_2} = \pm \frac{48 (n -1) n^2 \sqrt{(4 n +2) \left(k^2 (1-2 n )^2-16 (n -1) n \right)}}{k^3 (1-2 n)^4} .
\end{equation}
In the latter case, where the expression in the square root is negative and the point is a center, due to its positioning at the impasse surface, where time interchanges signs from one side to the other, the trajectories around the point start and end on the singular curve forming ``half-circles'' on each side, see Fig. \ref{S2center}.

Lastly, we want to comment on the stability properties of the impasse surface $ny+x^{2}=0$ as a whole. To deduce this, we follow the treatment presented in \cite{Saavedra2001,Alexandre2022}, where we write the system of equations (\ref{dynsysx}),(\ref{dynsysy}) in the following way
\begin{eqnarray}
    F_{ij}\dot z^{i}=E_j
    \label{eq:dynsysF}
\end{eqnarray}
where $\dot z^{i}=\begin{pmatrix}
        \dot x\\
        \dot y
    \end{pmatrix}$ is the velocity vector and
\begin{align}
    F_{ij}=\begin{pmatrix}
        x^2+ny & 0\\
        0 & x^2+ny
    \end{pmatrix}
\end{align}
\begin{align}
    E_j=\begin{pmatrix}
        \dfrac{3 x \left(y (x (k (3-2 n) n-k+n (2 n-1) x+x)+n (2 n-3))+(2 n-1) \left(x^2-1\right) x^2+n y^2\right)}{2 (2 n-1)}\\
        \dfrac{3 y \left(k \left(2 n^2-3 n+1\right) x^3+x^2 \left(\left(2 n^2-n+1\right) y-(1-2 n)^2\right)+(2 n-1) x^4+n (y-1) y\right)}{2 n-1} 
    \end{pmatrix}
\end{align}
Notice that $F\equiv\det F_{ij}=(x^2+ny)^2$, so that $F=0$ defines the impasse surface. Outside this surface one can invert $F_{ij}$ and solve (\ref{eq:dynsysF}) for the velocities,
\begin{eqnarray}
    \dot z^{i}=F^{ij}E_j ,
\end{eqnarray}
where $F^{ij}$ is the inverse of $F_{ij}$.

The stability properties of the impasse surface can be computed from the flux of the current $j^i=\sqrt{F}\dot z^i$ on the impasse surface. Let $n_i=\partial_i F$ be the normal vector field to the impasse surface, the flux is then given by
\begin{align}
    \Phi&=j^in_i=\sqrt{F}\,F^{ij}E_j\,\partial_i F\\
    &= \frac{3 (n -1)^2 x^4 \left(k (2 n -1) x-4 n\right)}{n (1-2n)} .
    \label{eq:fluxsys}
\end{align}
If $\Phi>0$, the orbits are directed away from the impasse surface, which therefore acts as a repulsor. If $\Phi<0$, the orbits are directed towards it, so it acts as an attractor. Points on the surface where the flux vanishes, $\Phi=0$, can be crossed by the orbits and may signal a boundary between an attracting and a repulsive sector of the impasse surface.

From \eqref{eq:fluxsys}, $\Phi$ vanishes at $x=0$ and at $x=x_{S_2}=\frac{4n}{(2n-1)k}$. Since $x=0$ is a root of even multiplicity (order four), $\Phi$ does not change sign there and this point does not separate different stability behaviors. The flux does change sign at $x=x_{S_2}$, which is precisely the location of $S_2$. Analyzing the sign of \eqref{eq:fluxsys} around this point shows that the impasse surface behaves as a repulsor when
\begin{equation}
    x< \frac{4 n}{(2 n - 1) k}, \qquad n\in\mathbb{R}^+-\left\{\tfrac{1}{2},1\right\},
    \label{surfrepcond1}
\end{equation}
or when
\begin{equation}
    x> \frac{4 n}{(2 n - 1) k}, \qquad n<0,
    \label{surfrepcond2}
\end{equation}
and it is an attractor otherwise.

The behavior corresponding to \eqref{surfrepcond2}, realized for $n<0$, is illustrated in Fig.~\ref{S2center}.

\subsection{Adding Radiation}

We conclude this section by exploring what properties of the cosmic evolution are modified by adding a radiation fluid. The Hamiltonian constraint of the model will acquire the following expression

\begin{equation}
    1-x^2-y-\Omega_m-\Omega_r=0
\end{equation}

Where $\Omega_r=\frac{\rho_r}{6f_0H^2}$, $\rho_r$ is the radiation energy density, that obeys the continuity equation $\dot\rho_r+4H\rho_r=0$, integration of this expression yields $\rho_r=\rho_{r,0}a^{-4}$. The effective equation of state parameter of the cosmic fluid is defined as

\begin{equation}
    w=-1-\frac{2}{3}\frac{\dot H}{H^2}=x^2+\frac{y}{2n-1}+\frac{\Omega_r}{3}
    \label{eq:EoSrad}
\end{equation}

The cosmological field equations expressed in the variables $x,y,\Omega_r$ are equivalent to

\begin{align}
x'={}&\frac{x}{2(2n-1)(x^2+ny)}
\Big(
3(2n-1)x^4-3k(n-1)(2n-1)xy
+x^2\big((3-3n+6n^2)y+(2n-1)(\Omega_r-3)\big) \notag\\
&\qquad\qquad\qquad\qquad
+ny\big(-9+6n+3y+(2n-1)\Omega_r\big)
\Big),\label{eq:dynxradmat}\\
y'={}&\frac{y\left(3(2n-1)x^2\left(1-2n+k(n-1)x+x^2\right)+3\left(-n+(1+n(2n-1))x^2\right)y+3ny^2r\right)}{(2n-1)(x^2+ny)} + y\Omega_r,\label{eq:dynyradmat}\\
\Omega_r'={}&\Omega_r\left(-1+3x^2+\frac{3y}{2n-1}+\Omega_r\right)\label{eq:dynOrradmat}.
\end{align}

Again, the system presents an impasse surface at $x^2+ny=0$; the properties of the surface as a whole are largely unmodified, since one can show that the flux acquires the same expression as (\ref{eq:fluxsys}).

The fixed points of the system (\ref{eq:dynxradmat}), (\ref{eq:dynyradmat}),(\ref{eq:dynOrradmat}), with the stability conditions, are presented in the table \ref{tablepoints_rad}, in Appendix \ref{appendix:fixed points radiation}. Generally, the properties of each fixed point are not substantially modified; the main difference is the appearance of the point $G$, where the $k$-essence field behaves as a radiation-like fluid while also having a non-vanishing radiation component. This point is neither stable nor unstable by any means; instead, it is a saddle (or saddle-focus) fixed point, so it cannot describe either early- or late-universe behavior; it always represents a transient cosmological phase. There are other fixed points that are inside the degeneracy surface $x^2+ny=0$, one of those points corresponds to a radiation-dominated solution $\Omega_r=1$, where the $k$-essence vanishes $x=y=0$.

With the dynamical properties described in the previous sections, one can use this model to describe early-universe behavior. To better understand this behavior, let's rewrite the system in terms of the variables $x, w, \Omega_r$.

\begin{align}
x'={}&\begin{aligned}[t]
\frac{x}{2}\Bigg[&\frac{\left(3w-3x^2-\Omega_r\right)\left(3(n-1)\left(x\left(2kn-k+2nx+x\right)-4n\right)+3(1-2n)nw+n(2n-1)\Omega_r\right)}{n(2n-1)\left(-3w+3x^2+\Omega_r\right)-3x^2}\\
&+3x^2+\Omega_r-3\Bigg],
\end{aligned}
\label{eq:dynx}\\
w'={}&\frac{3w\left((7-2n)n\Omega_r-9(n-1)x^2\left(2k(n-1)x-2n+1\right)\right)+6(n-1)^2x^2\Omega_r(3kx-2)+54(n-1)^2x^4(kx-1)}{3n(2n-1)\left(3w-3x^2-\Omega_r\right)+9x^2}
\notag\\
&\qquad+\frac{-9w^2\left(3(n-1)(2n+1)x^2+n\left((2n-1)\Omega_r+3\right)\right)+27n(2n-1)w^3+2(n-2)n\Omega_r^2}{3n(2n-1)\left(3w-3x^2-\Omega_r\right)+9x^2},
\label{eq:dynw}\\
\Omega_r'={}&(3w-1)\Omega_r.
\label{eq:dynOrrad}
\end{align}

One can start from an inflationary initial condition, provided by an unstable degeneracy surface, followed by the evolution toward point~$D$. Subsequently, the $k$-essence field evolves into an (almost) radiation-like state before finally settling into point~$F$, where it behaves as a dark matter component with energy density parameters bounded by $\Omega_i \in (0,1)$, where $i$ denotes a specific energy component. This behavior is realized for parameters satisfying $0 < k < \sqrt{3}$ and $1 - \frac{k^2}{2} < n \le \frac{7}{2} + \frac{20}{k^2 - 8}$, which corresponds to the admissible parameter range for interpreting the $k$-essence field as dark matter. Almost any initial condition fulfilling $x(0) > 0$ and $w(0) < -\frac{1}{3}$ can reproduce this behavior in a physically viable manner at the background level.

\begin{figure}
    \centering
    \includegraphics[width=0.6\linewidth]{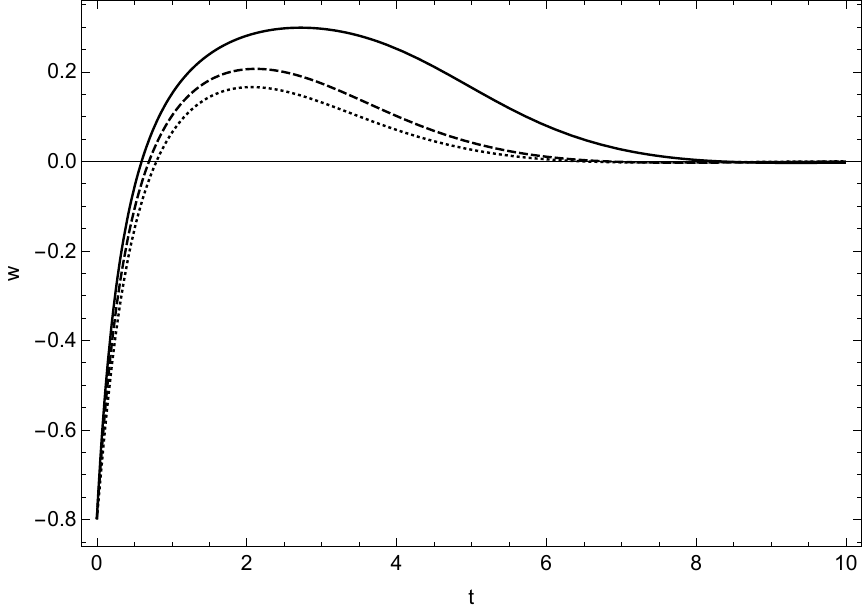}
    \caption{Numerical integration of the effective equation of state $w(t)$ for three distinct initial conditions, considering $n=3/4$ and $k=1.1$. All trajectories start at $w(0) = -0.8$ and $\Omega_r(0) = 0.3$. The solid, dashed, and dotted curves correspond to $x(0) = 1.5$, $x(0) = 1.4$, and $x(0) = 1.3$, respectively.}
    \label{fig:num_sol_rad}
\end{figure}

\begin{figure}
    \centering
    \includegraphics[width=0.65\linewidth]{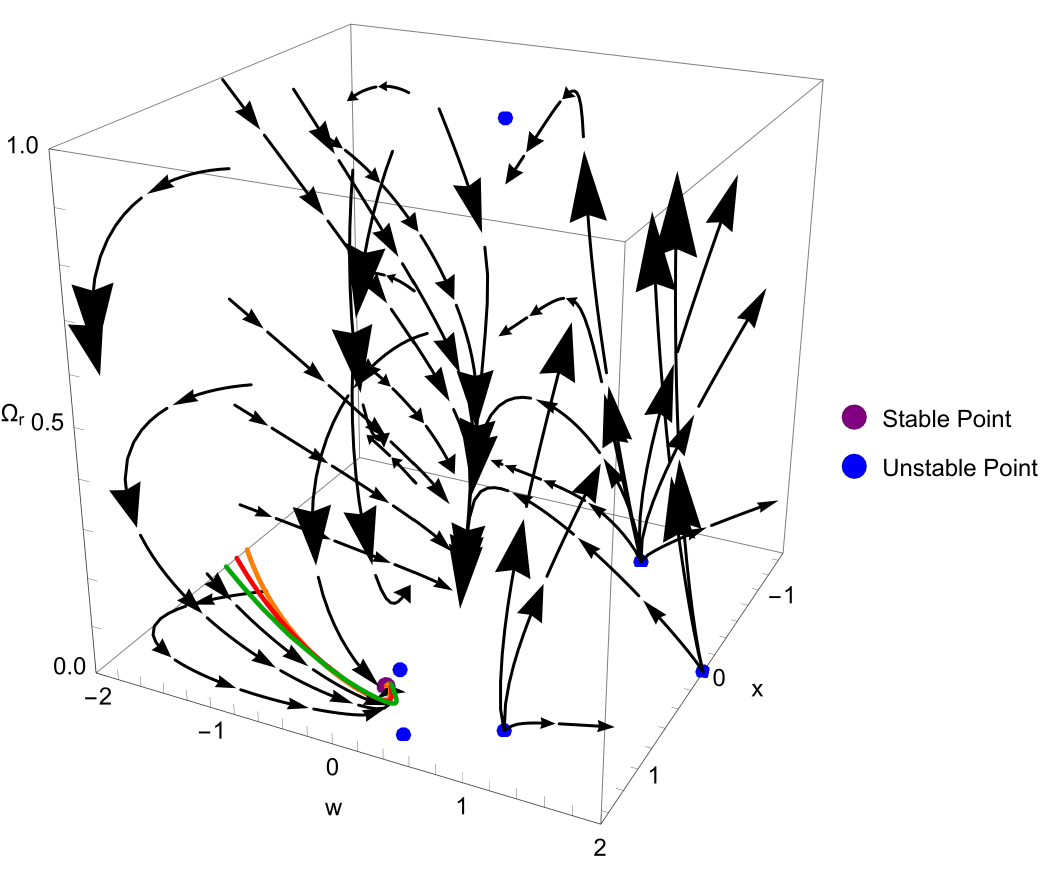}
    \caption{Phase space portrait of the system for $n=3/4$ and $k=1.1$. Unstable critical points ($A, B, C, D, E,$ and $G$), as well as the pure radiation state, are depicted in blue. The only stable fixed point, $F$ (purple), corresponds to a state where the $k$-essence field mimics pressureless matter. The green, red, and orange curves denote trajectories initialized at $w(0)=-0.8$ and $\Omega_r(0)=0.3$, with $x(0)=1.5$, $x(0)=1.4$, and $x(0)=1.3$, respectively, the initial conditions are chosen such that $\Omega_m\in(0,1)$. The orbits naturally converge towards the attractor $F$.}
    \label{fig:trayectories_radiation}
\end{figure}

The effect of choosing different initial conditions on the evolution of the effective equation of state is illustrated in Fig.~\ref{fig:num_sol_rad}, while the numerical integration of the corresponding orbits is shown in Fig. \ref{fig:trayectories_radiation}. However, as seen in Fig.~\ref{fig:poinc_disc}, an initial condition chosen too close to the impasse surface can yield negative energy densities for the $k$-essence field; therefore, the initial conditions must be constrained to ensure a positive initial energy density.

\section{Conclusions}
\label{sec6}
In this work, we investigated a $k$-essence cosmological model in a spatially flat FLRW background in which the free functional form of the theory was constrained by a geometric selection rule, rather than by an ad hoc ansatz. Specifically, we considered the generic $k$-essence Lagrangian $F(R,X,\phi)=f_{1}R+f_{2}(\phi,X)$ and we required the gravitational field equations to possess a nontrivial variational symmetry, that is, there exists a transformation such that the action integral is transformed and the variation remains invariant. These cosmological field equations possess a nontrivial variational symmetry and a conservation law, which reduce the order of the field equations and constrain the trajectories in the phase-space, thereby characterizing the model as Liouville integrable.

To explore the phase space, we reformulated the field equations in terms of dimensionless variables within the Hubble-normalization framework. For the new dimensionless dynamical system, we determined the stationary points, and we studied their stability properties in three cases: in the absence of a matter source, with a nonzero matter term, and with a nonzero radiation term. 

In the matter-free case, the asymptotic states in the finite regime describe stiff fluid solutions  (points $A$ and $B$), scaling solutions with an equation of state parameter $w=-1\mp k\Gamma_{\pm}$ (points $C$ and $D$), and an additional one at point $E$. In the presence of matter, an extra fixed point, $F$, emerges, in which the $k$-essence field behaves like a pressureless matter component, which contributes to the dark matter. As a result, the model can describe the early-universe phenomenology; an initial unstable inflationary epoch, and its subsequent transition to a matter-dominated era. To complement our analysis, we apply a Poincar\'{e} transformation to compactify the phase space and explore the asymptotic behavior of trajectories at infinity. The resulting points correspond to a diverging equation of state parameter. {This constrains both the initial conditions of this model and the sign of the parameter $\beta$. Even if all energy densities are initially positive, the model can evolve into a regime where the energy density of the $k$-essence field becomes negative. Consequently, not every set of initial conditions with positive energy densities is allowed.}

Finally, we explore what occurs in the presence of radiation and of a nonzero mass term for the $k$-essence field. When we introduce radiation to the model, we found the existence two new families of solutions, which describe radiation dominated universes. On the other hand, for the case of an exponential potential the admitted asymptotic solutions have a similar structure with the massless field and quintessence with exponential potential, at least for the particular choice of values presented in \ref{append2}, thus, no new significantly different families of solution trajectories are provided. 

The symmetry-driven $k$-essence cosmological model is a minimal model that possesses a variational symmetry which constrains the solution space. In a future study, we plan to employ observational data to constrain the free parameters of the model, as well as to explore the application of the symmetry analysis to a more general form of $k$-essence theory. Finally, the existence of the variational symmetry provides the quantum operators required for the minisuperspace quantization of the model, connecting the present classical analysis to the canonical quantum description developed in \cite{Lueiza2026}. 
We plan to address these questions in future work.

\begin{acknowledgments}
GL \& AP acknowledges the support from FONDECYT Grant 1240514. GL \& AP thank the Universidad de La Frontera for the hospitality extended during part of this work. AL acknowledges financial support from Universidad de La Frontera. GL thanks the support of Vicerrectoría de Investigación y Desarrollo Tecnológico (VRIDT) of Universidad Católica del Norte (UCN) through Resoluci\'{o}n VRIDT No. 096/2022, Resoluci\'{o}n VRIDT No. 200/2025, and Resoluci\'{o}n VRIDT No. 021/2026. AP acknowledges the COST Action CA23130 ``Bridging high and low energies in search of quantum gravity (BridgeQG)''
\end{acknowledgments}

\appendix

\section{Fixed Points of the Model with Pressureless Matter and Radiation}\label{appendix:fixed points radiation}

\begin{longtable}{c|c|c|c|p{4.5cm}|p{3.5cm}}
\caption{Fixed Points and Stability Conditions for the Cosmological System Including Radiation}
\label{tablepoints_rad}
\renewcommand{\arraystretch}{1.3} 
\\
\hline\hline
 \textbf{Point $(x,y,\Omega_r)$} & $\boldsymbol{\Omega_m}$ & \textbf{$w$} & \textbf{Classification} & $\boldsymbol{\ n}$ & $\boldsymbol{\ k}$ \\
\hline\hline

\multirow{2}{*}{A: $(-1,0,0)$} & \multirow{2}{*}{$0$} & \multirow{2}{*}{$1$}
    & Saddle & $n > 1$ & $k>0$ \\
\cline{4-6}
    & & & Repulsor & $n < 1$ & $k>0$ \\
\hline

\multirow{3}{*}{B: $(1,0,0)$} & \multirow{3}{*}{$0$} & \multirow{3}{*}{$1$}
    & \multirow{2}{*}{Saddle} & $n < 1$ & $k > 2$ \\
\cline{5-6}
    & & & & $n > 1$ & $0 < k < 2$ \\
\cline{4-6}
    & & & Repulsor & \multicolumn{2}{c}{Otherwise} \\
\hline

\multirow{11}{*}{\shortstack{C:\\$(\Gamma_+,\, \left(\frac{2}{k}-\Gamma_+)\delta,0\right)$}} & \multirow{11}{*}{$0$} & \multirow{11}{*}{$-1 + k\Gamma_+$}
    & \multirow{3}{*}{Attractor} & $0<n<1$ & $k > 0$ \\
\cline{5-6}
    & & & & $-\frac{1}{2} < n < 0$ & $k>4\frac{\sqrt{(-1+n)n}}{1-2n}$ \\
\cline{5-6}
    & & & & $n\le -\frac{1}{2}$ & $k>\sqrt{2-2n}$ \\
\cline{4-6}
    & & & \multirow{6}{*}{Saddle} & \rule{0pt}{15pt} $-1<n<-\frac{1}{2}$ & $4\frac{\sqrt{(-1+n)n}}{1-2n}<k<\sqrt{2-2n}$\\
\cline{5-6}
    & & & & \rule{0pt}{15pt} $n\le-1$ & $4\sqrt{\frac{n-1}{3(n-2)}}<k<\sqrt{2-2n}$ \\
\cline{5-6}
& & & & \rule{0pt}{15pt} $n<-1$ & $4\frac{\sqrt{(-1+n)n}}{1-2n}<k<4\sqrt{\frac{n-1}{3(n-2)}}$ \\
\cline{4-6}
   & & & \multirow{1}{*}{Repulsor} & \rule{0pt}{15pt} $n>1$ & \rule{0pt}{16pt} $k>4\frac{\sqrt{(-1+n)n}}{-1+2n}$ \\
   \cline{5-6}

\hline

\multirow{10}{*}{\shortstack{D:\\$\left(\Gamma_-,\, (\frac{2}{k}-\Gamma_-)\delta,0\right)$}} & \multirow{10}{*}{$0$} & \multirow{10}{*}{$-1 + k\Gamma_-$}
    & \multirow{1}{*}{Attractor} & $\frac{1}{2} -\frac{1}{\sqrt{4-k^2}}< n <1-\frac{k^2}{2}$ & $0<k<\sqrt{3}$ \\
\cline{4-6}
    & & & \multirow{7}{*}{Saddle} & $n>2+\frac{16}{3k^2-16}$ & $k> \frac{4}{\sqrt{3}}$ \\
\cline{5-6}
& &&& $1-\frac{k^2}{2} < n < 2+\frac{16}{3k^2-16}$ & $0< k \le \sqrt{3}$ \\
\cline{5-6}
    & & & & $\frac{1}{2} -\frac{1}{\sqrt{4-k^2}}<n< 2+\frac{16}{3k^2-16}$ & $\sqrt{3} < k <\frac{4\sqrt{2}}{3}$ \\
\cline{5-6}
    & & & & $2+\frac{16}{3k^2-16}<n<1$ & $0<k \le \frac{4\sqrt{2}}{3}$ \\
    \cline{5-6}
    & & & & $\frac{1}{2} -\frac{1}{\sqrt{4-k^2}}<n<1$ & $\frac{4\sqrt{2}}{3}<k<2$ \\
    \cline{5-6}
    & & & & $n>1$ & $2<k\le\frac{4}{\sqrt{3}}$ \\
    \cline{5-6}
    & & & & $1<n<2+\frac{16}{3k^2-16}$ & $k>\frac{4}{\sqrt{3}}$ \\
\cline{4-6}
    & & & \multirow{2}{*}{Repulsor} & $1<n<\frac{1}{2}+\frac{1}{\sqrt{4-k^2}}$ & $0<k<2$ \\
\cline{5-6}
    & & & & $n < 1$ & $k>2$ \\
\hline

\multirow{2}{*}{E: $(0,1,0)$} & \multirow{2}{*}{$0$} & \multirow{2}{*}{$\dfrac{1}{2n-1}$}
    & Repulsor & $1<n < 2$ & $k>0$ \\

\cline{4-6}
     & & & Saddle & \multicolumn{2}{c}{Otherwise} \\
\hline

\multirow{7}{*}{\shortstack{F:\\$(\frac{1}{k},\, \frac{1-2n}{k^2},0)$}} & \multirow{7}{*}{$1 + \dfrac{2(n-1)}{k^2}$} & \multirow{7}{*}{$0$}
    & \multirow{3}{*}{Attractor} & $1-\frac{k^2}{2} < n \leq  \frac{7}{2}+\frac{20}{k^2-8}$ & $0 < k < \sqrt{3}$ \\
\cline{5-6}
    & & & & $ \frac{7}{2}+\frac{20}{k^2-8} \le n<1-\frac{k^2}{2}$ & $\sqrt{3}<k<2\sqrt{2}$ \\
\cline{5-6}
    & & & & $ n\ge \frac{7}{2}+\frac{20}{k^2-8}\quad\text{or}\quad n<1-\frac{k^2}{2}$ & $k>2\sqrt{2}$ \\
\cline{4-6}
    & & & \multirow{1}{*}{Saddle} & $-\frac{1}{2}< n < 1-\frac{k^2}{2}$ & $0<k<\sqrt{3}$ \\
\cline{4-6}
    & & & \multirow{3}{*}{Stable Spiral} & $n<-\frac{1}{2}\quad \text{or}\quad n \ge \frac{7}{2}+\frac{20}{k^2-8}$ & $0 < k < \sqrt{3}$ \\
\cline{5-6}
    & & & & $n\le \frac{7}{2}+\frac{20}{k^2-8} \quad \text{or}\quad n>-\frac{1}{2}$ & $\sqrt{3}<k<2\sqrt{2}$ \\
\cline{5-6}
    & & & & $-\frac{1}{2}<n< \frac{7}{2}+\frac{20}{k^2-8}$ & $k>2\sqrt{2}$ \\

\hline

\multirow{7}{*}{\shortstack{G:\\$(\frac{4}{3k},\, \frac{16(-1+2n)}{9k^2(-2+n)},1 - \frac{16(n-1)}{3k^2(n-2)})$}} & \multirow{7}{*}{$0$} & \multirow{7}{*}{$\dfrac{1}{3}$}
    & \multirow{2}{*}{Saddle} & $-1 < n < 2+\frac{16}{3k^2-16}$ & $0 < k < 4\frac{\sqrt{2}}
    {3}$ \\
\cline{5-6}
    & & & & $2+\frac{16}{3k^2-16} < n $ & $\frac{4}{\sqrt{3}}<k$ \\

\cline{5-6}
    & & & & $n<-1 $ & $\frac{4}{\sqrt{3}}\le k$ \\
\cline{5-6}
    & & & & $2+\frac{16}{3k^2-16}<n<-1 $ & $4\frac{\sqrt{2}}{3}< k<\frac{4}{\sqrt{3}}$ \\
\cline{4-6}
    & & & \multirow{3}{*}{Saddle Focus} & $n<-1\quad \text{or}\quad n \ge \frac{45k^2-128}{27k^2-128}$ & $0 < k\le 4\frac{\sqrt{2}}{3}$ \\
\cline{5-6}
    & & & & $n\le\frac{45k^2-128}{27k^2-128}\quad \text{or} \quad n>-1$ & $4\frac{\sqrt{2}}{3}< k < 8\frac{\sqrt{6}}{9}$ \\
\cline{5-6}
    & & & & $-1<n< \frac{45k^2-128}{27k^2-128}$ & $k>8\frac{\sqrt{6}}{9}$ \\
\hline

\end{longtable}

\section{Model With Exponential Potential}
\label{append2}

We continue with the introduction of a nonzero mass term, described by the exponential potential such that the scaling symmetry to be preserved. In particular we consider the $k$-essence theory with

\begin{equation}
f_{1}(\phi )=f_{0},\qquad f_{2}(\phi,X)=\alpha X-V_0\exp(-\phi)+\beta \exp \left(
(n-1)\phi \right) X^{n},  \label{mod1_pot}
\end{equation}%

The field equations of this given model can be expressed in terms of the dimensionless variables $x,y$ in pressence of a pressureless matter source as follows

\begin{align}
x' &= \frac{3}{2} x \left( 1 + x^2 + \frac{y}{2n-1} - c \left(x^{-2n} y\right)^{\frac{1}{1-n}} + \frac{(2-4n)x^2 - 2ny - k(2n-1)x \left((n-1)y - c \left(x^{-2n} y\right)^{\frac{1}{1-n}}\right)}{(2n-1)(x^2 + ny)} \right),\label{eq:sysxpot} \\[1.5ex]
y' &= \frac{3y}{(2n-1)(x^2 + ny)} \left[ (2n-1)^2 x^2 + k(n-1)(2n-1) x^3 + (2n-1) x^4 - ny + (2n^2 - n + 1) x^2 y + ny^2 \right. \notag\\
   &\quad \left.  - c(2n-1)\left(x^{-2n} y\right)^{\frac{1}{1-n}} \left(x^2 + ny - kn x\right) \right]\label{eq:sysypot},
\end{align}
where $c = V_0 (2n-1)^{\frac{1}{n-1}} \alpha^{-\frac{n}{n-1}} \beta^{\frac{1}{n-1}}$. Furthermore, the effective equation of state parameter is given by the expression
\begin{eqnarray}
    w=x^2 + \frac{y}{2n-1} - c \left(x^{-2n} y\right)^{\frac{1}{1-m}} ,
\end{eqnarray}
where the Hamiltonian constraint reads

\begin{eqnarray}
    1 - x^2 - y - c \left(x^{-2n} y\right)^{\frac{1}{1-n}} - \Omega_m = 0 .
\end{eqnarray}

\begin{figure}
    \centering
    \includegraphics[width=0.65\linewidth]{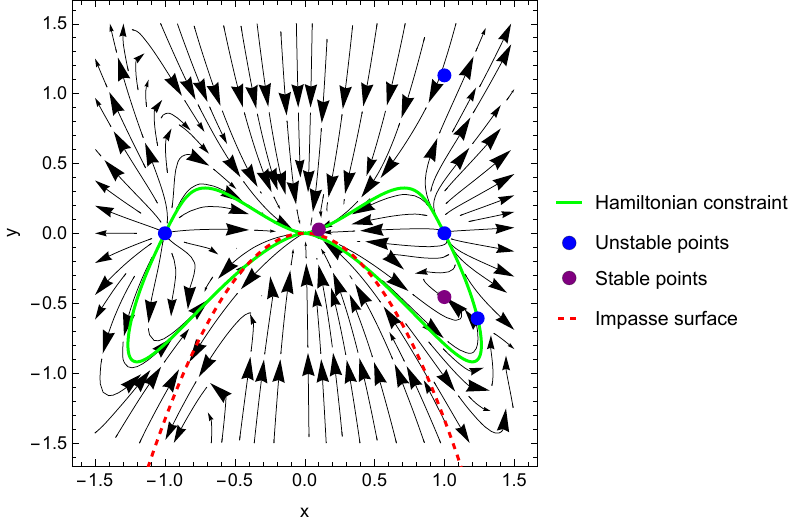}
    \caption{Phase-space portrait of the model (\ref{mod1_pot}), unstable points ($A,B,C,D$) are drawn in blue, stable points $(E,F)$ are drawn in purple. Similar to the model without potential, there is an impasse surface at $x^2+ny=0$, which, for $n=3/4$, is shown as a red dashed line.}
    \label{fig:phase_space_pot}
\end{figure}

Due to the $n$-dependence of the exponents in the dynamical equations, any analytical treatment of the fixed points is rendered impossible in the variables $x,y$, so we will only focus on specific values of the parameters and a numerical analysis of the fixed points. For $n=3/4$ and $k=c=1$ the fixed points are presented in the Table \ref{tablepoints_pot}, while the phase space portrait is shown in Fig. \ref{fig:phase_space_pot}. The phase space portrait clearly shows that the system can be used to model the early universe behavior, connecting inflation to dark-matter, which is the result already mentioned in this paper, or to model the transition from a matter dominate epoch to a late-time accelerated Universe, where the $k$-essence effectively plays the role of dark-energy, this behavior can already achieved with pure quintessence. We conclude that adding an exponential mass term, the cosmological dynamics provide the same cosmological history and evolution for the values of interest.

\begin{table}[htbp]
\centering
\caption{Fixed Points and Stability Conditions of the system (\ref{eq:sysxpot}),(\ref{eq:sysypot})}
\renewcommand{\arraystretch}{1.3}

\label{tablepoints_pot}
\begin{tabular}{c|c|c|c}
\hline\hline
\textbf{Point $(x,y)$} & $\boldsymbol{\Omega_m}$ & \textbf{$w$} & \textbf{Classification} \\
\hline
A: $(-1,0)$ & $0$ & $1$ & Repulsor \\
\hline
B: $(1,0)$ & $0$ & $1$ & Repulsor \\
\hline
C: $(1.2380, -0.6091)$ & $0.2380$ & $0$ & Saddle \\
\hline
D: $(1, 1.1299)$ & $-4.3897$ & $0$ & Saddle \\
\hline
E: $(1, -0.4566)$ & $0.3697$ & $0$ & Attractor \\
\hline
F: $(0.1017, 0.0270)$ & $0$ & $-0.8983$ & Attractor \\
\hline
\end{tabular}
\end{table}

\end{document}